\documentclass[preprint,aps,12pt,showpacs,tightenlines]{revtex4}
\usepackage{amsmath}
\usepackage{amssymb}
\usepackage{epsf,epsfig,graphicx}
\begin{document}
\preprint{NJNU-TH-06-25}

\newcommand{\beq}{\begin{eqnarray}}
\newcommand{\eeq}{\end{eqnarray}}
\newcommand{\non}{\nonumber\\ }
\newcommand{\etap}{\eta^{(\prime)} }
\newcommand{\ov  }{\overline  }
\newcommand{\acp}{{\cal A}_{CP}}

\newcommand{\psl}{ p\hspace{-1.8truemm}/ }
\newcommand{\nsl}{ n\hspace{-2.2truemm}/ }
\newcommand{\vsl}{ v\hspace{-2.2truemm}/ }
\newcommand{\epsl}{\epsilon\hspace{-1.8truemm}/\,  }

\def \cpl{ Chin. Phys. Lett.  }
\def \ctp{ Commun. Theor. Phys.  }
\def \epjc{ Eur. Phys. J. C }
\def \jhep{ J. High Energy Phys.  }

\def \jpg{  J. Phys. G }
\def \npb{  Nucl. Phys. B }
\def \plb{  Phys. Lett. B }
\def \prd{  Phys. Rev. D }
\def \prl{  Phys. Rev. Lett.  }
\def \zpc{  Z. Phys. C  }

\title{$B \to K \; K^{*}$ Decays in the Perturbative QCD Approach}
\author{Libo  Guo } \email{guolibo@njnu.edu.cn}
\author{Qian-gui Xu}
\author{Zhen-jun Xiao} \email{xiaozhenjun@njnu.edu.cn}
\affiliation{Department of Physics and Institute of Theoretical Physics, Nanjing Normal
University, Nanjing, Jiangsu 210097, P.R.China}
\date{\today}
\begin{abstract}
We calculate the branching ratios and CP-violating asymmetries for
$B^0 \to K^{0} \ov K^{*0}$, $\ov K^{0} K^{*0}$, $K^+ K^{*-}$,
$K^- K^{*+}$, and $B^+\to K^+ \ov K^{*0}$ and $ \ov K^0 K^{*+} $
decays by employing the low energy effective Hamiltonian and the
perturbative QCD (pQCD) factorization approach. The theoretical
predictions for the branching ratios are $Br(B^0/\ov B^0 \to K^{\pm}
K^{*\mp}) \approx 7.4 \times 10^{-8}$ , $Br(B^0/\ov B^0 \to K^{0}
\ov K^{*0}(\ov K^{0} K^{*0})) \approx 19.6 \times 10^{-7}$ ,
$Br(B^+\to K^+ \ov K^{*0}) \approx 3 \times 10^{-7}$ and $Br(B^+\to
K^{*+} \ov K^0) \approx 18.3 \times 10^{-7}$, which are consistent
with currently available experimental upper limits. We also predict
large CP-violating asymmetries in these decays: $A_{CP}^{dir}(K^\pm
\ov K^{*0})\approx -20 \%$, $A_{CP}^{dir}(K^{*\pm} \ov K^0)\approx
-49\%$, which can be tested by the forthcoming B meson experiments.
\end{abstract}
\pacs{13.25.Hw, 12.38.Bx, 14.40.Aq, 14.40.Ev}

\maketitle

\section{Introduction}

The study of exclusive non-leptonic weak decays of B mesons provides not only good opportunities for
testing the Standard Model (SM) but also powerful means for probing different new physics scenarios
beyond the SM. The mechanism of two body B decay is still not quite clear, although many scientists
devote to this field. Starting from factorization hypothesis \cite{fact}, many approaches have been built
to explain the existing data and some progresses have been made. For example the generalized
factorization (GF)\cite{ali98}, QCD factorization (QCDF) approach \cite{du02,bn03b}, the
perturbative QCD (pQCD) approach \cite{lb80,pqcd,kls2001,li2003} and the soft-collinear
effective theory (SCET)
\cite{scet}. The pQCD approach is based on $K_T$ factorization theorem\cite{theorem} while
others are mostly based on collinear factorization \cite{coll}.

In our opinion, the pQCD factorization approach has three special features:
(a) Sudakov factor and threshold resummation \cite{resum} are included to regulate the
end-point singularities, so the arbitrary cutoff\cite{cut} is no longer necessary;
(b) the form factors for $B \to M$ transition can be calculated perturbatively, although
some controversies still exist about this point;
and (c) the annihilation diagrams are calculable and play an important role in producing
CP violation \cite{li2003,hsw06}.
Up to now, many B meson decay channels have been studied by employing the pQCD approach,
and it has become one of the most popular methods to calculate the hadronic matrix elements.

In this paper, we will study the branching ratios and CP asymmetries of $B \to K\; K^*$ decays
in the pQCD factorization approach. Theoretically, in the $B\to K K^*$ decay modes, the B meson is
heavy and sitting at rest. It decays into two light mesons with large momenta, so
these two energetic final state mesons may have no enough time to get involved in soft
final state interaction (FSI).
In this case, the short distance hard process dominates the decay amplitude and the non-perturbative
FSI effects may not be important, this makes the pQCD approach applicable.
At the same time, the $B \to K\;K^*$ decays have been studied before in
the GF approach \cite{ali98} and the QCDF approach \cite{du02,bn03b}.
The similar decays such as $B \to K K$ and
$K^* K^*$ decays have been investigated in the pQCD approach recently \cite{chen00a,zhu05a}.
On the experimental side, the first measurement of $B^0 \to ( K^0 \ov K^{*0}+
 \ov K^0 K^{*0}) $ decay has been reported very recently by BaBar collaboration \cite{babar06}
in units of $10^{-6}$ (upper limits at $90\%$ C.L.):
\beq
Br(B^0 \to K^0 \ov K^{*0}+  \bar K^0 K^{*0} )&=&  0.2 ^{+0.9+0.1}_{-0.8-0.3}\ \ (< 1.9).
\label{eq:exp}
\eeq
For $B^+ \to K^+ \ov K^{*0}$ decay, only the experimental upper limit is available
now \cite{pdg2006,hfag06}
\beq
Br(B^+ \to K^+ \ov K^{*0})< 5.3 \times 10^{-6}.
\label{eq:ulimits}
\eeq

This paper is organized as follows. In Sec. II, we give the theoretical framework of the
pQCD factorization approach. Next, we calculate the relevant Feynman diagrams and present
the various decay amplitudes for $B\to K K^*$ decays.
In Sec. IV, we show the numerical results of the CP-averaged branching ratios and CP asymmetries
and compare them with currently available experimental measurements or the theoretical predictions
in QCDF approach. The summary and some discussions are included in the final section.

\section{ Theoretical framework}\label{sec:f-work}

\begin{figure}[tb]
\vspace{-.6 cm}
 \centerline{\epsfxsize=10 cm \epsffile{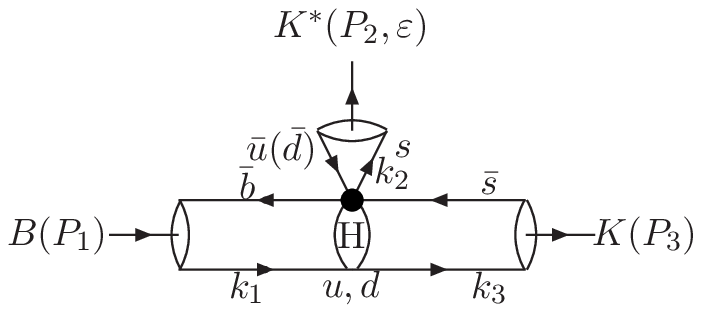}}
\caption{Factorization for $B\to K K^*$ Decays }
 \label{fig:fig1}
\end{figure}

The three scales pQCD factorization approach\cite{pqcd,kls2001} has been developed and applied in the
non-leptonic $B$ meson decays for some time. In this approach, the decay amplitude is
factorized into the convolution of the mesons'
light-cone wave functions, the hard scattering kernel and the Wilson coefficients,
as illustrated schematically by Fig.~1, which stands for the soft, hard
and harder dynamics characterized by three different energy scales
$(t\sim {\cal O}(\sqrt{\overline{\Lambda}M_B}), m_b, M_W)$ respectively.
Then the decay amplitude ${\cal A}(B \to M_1 M_2)$ is conceptually written as the convolution
\beq
{\cal A}(B \to M_1 M_2)\sim \int\!\! d^4k_1 d^4k_2 d^4k_3\ \mathrm{Tr}
\left [ C(t) \Phi_B(k_1) \Phi_{M_1}(k_2) \Phi_{M_2}(k_3) H(k_1,k_2,k_3, t)
\right ],
\label{eq:con1}
\eeq
where $k_i$'s are momenta of light quarks included in each mesons,
and the term $``\mathrm{Tr}"$ denotes the trace over Dirac and color indices.
$C(t)$ is the Wilson coefficient which results from the radiative
corrections at short distance. In the above convolution, $C(t)$
includes the harder dynamics at scale larger than $M_B$ and
describes the evolution of local $4$-Fermi operators from $m_W$ (the
$W$ boson mass) down to $t\sim\mathcal{O}(\sqrt{\bar{\Lambda} M_B})$
scale, where $\bar{\Lambda}\equiv M_B -m_b$. The function
$H(k_1,k_2,k_3,t)$ describes the four quark operator and the
spectator quark connected by a hard gluon whose $q^2$ is of the
order of $\bar{\Lambda} M_B$, and includes the
$\mathcal{O}(\sqrt{\bar{\Lambda} M_B})$ hard dynamics. Therefore,
this hard part $H$ can be evaluated as an enpansion in power of
$\alpha_S(t)$ and $\bar{\Lambda}/t$, and depends on the processes considered.
The function $\Phi_M$ ($M=B, M_1, M_2$) is the wave function which describes
hadronization of the quark and anti-quark into the meson $M$, and independent
of the specific processes. Using the wave functions
determined from other well measured processes, one can make quantitative predictions here.

Since the b quark is rather heavy we consider the $B$ meson at rest
for simplicity. It is convenient to use light-cone coordinate $(p^+,
p^-, {\bf p}_T)$ to describe the meson's momenta,
\beq
p^\pm =\frac{1}{\sqrt{2}} (p^0 \pm p^3), \quad  and \quad {\bf p}_T = (p^1,
p^2).
\eeq
Using the light-cone coordinates the $B$ meson and the two final state meson
momenta can be written as
\beq
P_1 = \frac{M_B}{\sqrt{2}} (1,1,{\bf 0}_T),
\quad P_2 = \frac{M_B}{\sqrt{2}}(1,r_{k^*}^2,{\bf 0}_T),
\quad P_3 = \frac{M_B}{\sqrt{2}} (0,1-r_{k^*}^2,{\bf 0}_T),
\eeq
respectively, where $r_{K^*}=m_{K^*}/m_B$; and the terms proportional to $m_K^2/m_B^2$
have been neglected.

For the $B \to K K^*$ decays considered here, only the $K^*$ meson's
longitudinal part contributes to the decays, its polarization vector is
$\epsilon_L=\frac{M_B}{\sqrt{2}M_{K^*}} (1,-r_{K^{*}}^2,\bf{0_T})$.
Putting the light (anti-) quark momenta in $B$, $K^*$ and $K$ mesons
as $k_1$, $k_2$, and $k_3$, respectively, we can choose
\beq k_1 =
(x_1 P_1^+,0,{\bf k}_{1T}), \quad k_2 = (x_2 P_2^+,0,{\bf k}_{2T}),
\quad k_3 = (0, x_3 P_3^-,{\bf k}_{3T}).
\eeq
Then the integration
over $k_1^-$, $k_2^-$, and $k_3^+$ in Eq.(\ref{eq:con1}) will lead to
\beq
{\cal A}(B \to K K^*) &\sim &\int\!\! d x_1 d x_2 d x_3 b_1
d b_1 b_2 d b_2 b_3 d b_3 \non && \quad \mathrm{Tr} \left [ C(t)
\Phi_B(x_1,b_1) \Phi_{k^*}(x_2,b_2) \Phi_k(x_3, b_3) H(x_i, b_i, t)
S_t(x_i)\, e^{-S(t)} \right ], \label{eq:a2}
\eeq
where $b_i$ is the conjugate space coordinate of $k_{iT}$, and $t$ is the largest
energy scale in function $H(x_i,b_i,t)$. The large logarithms $\ln(m_W/t)$
coming from QCD radiative corrections to four quark
operators are included in the Wilson coefficients $C(t)$. The large
double logarithms ($\ln^2 x_i$) on the longitudinal direction are
summed by the threshold resummation\cite{resum}, and they lead to
$S_t(x_i)$ which smears the end-point singularities on $x_i$. The
last term, $e^{-S(t)}$, is the Sudakov form factor resulting from
overlap of soft and collinear divergences, which suppresses the soft
dynamics effectively\cite{soft}. Thus it makes the perturbative
calculation of the hard part $H$ applicable at intermediate scale,
i.e., $M_B$ scale.


The weak effective Hamiltonian $H_{eff}$ for $B \to K K^*$ decays can be written as \cite{buras96}
\beq
\label{eq:heff} {\cal H}_{eff} = \frac{G_{F}}
{\sqrt{2}} \, \left[ V_{ub} V_{ud}^* \left (C_1(\mu) O_1^u(\mu) +
C_2(\mu) O_2^u(\mu) \right) - V_{tb} V_{td}^* \, \sum_{i=3}^{10}
C_{i}(\mu) \,O_i(\mu) \right] \; .
\eeq
where $C_i(\mu)$ are Wilson coefficients evaluated at the renormalization scale $\mu$ and $O_i$
are the four-fermion operators for $b \to d$ transition:
\beq
\begin{array}{llllll}
O_1^{u} & = &  \bar d_\alpha\gamma^\mu L u_\beta\cdot \bar
u_\beta\gamma_\mu L b_\alpha\ , &O_2^{u} & = &\bar
d_\alpha\gamma^\mu L u_\alpha\cdot \bar
u_\beta\gamma_\mu L b_\beta\ , \\
O_3 & = & \bar d_\alpha\gamma^\mu L b_\alpha\cdot \sum_{q'}\bar
 q_\beta'\gamma_\mu L q_\beta'\ ,   &
O_4 & = & \bar d_\alpha\gamma^\mu L b_\beta\cdot \sum_{q'}\bar
q_\beta'\gamma_\mu L q_\alpha'\ , \\
O_5 & = & \bar d_\alpha\gamma^\mu L b_\alpha\cdot \sum_{q'}\bar
q_\beta'\gamma_\mu R q_\beta'\ ,   & O_6 & = & \bar
d_\alpha\gamma^\mu L b_\beta\cdot \sum_{q'}\bar
q_\beta'\gamma_\mu R q_\alpha'\ , \\
O_7 & = & \frac{3}{2}\bar d_\alpha\gamma^\mu L b_\alpha\cdot
\sum_{q'}e_{q'}\bar q_\beta'\gamma_\mu R q_\beta'\ ,   & O_8 & = &
\frac{3}{2}\bar d_\alpha\gamma^\mu L b_\beta\cdot
\sum_{q'}e_{q'}\bar q_\beta'\gamma_\mu R q_\alpha'\ , \\
O_9 & = & \frac{3}{2}\bar d_\alpha\gamma^\mu L b_\alpha\cdot
\sum_{q'}e_{q'}\bar q_\beta'\gamma_\mu L q_\beta'\ ,   & O_{10} &
= & \frac{3}{2}\bar d_\alpha\gamma^\mu L b_\beta\cdot
\sum_{q'}e_{q'}\bar q_\beta'\gamma_\mu L q_\alpha'\ ,
\label{eq:operators}
\end{array}
\eeq
where $\alpha$ and $\beta$ are the $SU(3)$ color indices; $L$
and $R$ are the left- and right-handed projection operators with
$L=(1 - \gamma_5)$, $R= (1 + \gamma_5)$. The sum over $q'$ runs
over the quark fields that are active at the scale $\mu=O(m_b)$,
i.e., $(q'\epsilon\{u,d,s,c,b\})$. For the decays with $b \to s$ transition, simply
make a replacement of $d$ by $s$ in Eqs.~(\ref{eq:heff}) and (\ref{eq:operators}).

The pQCD approach works well for the leading twist approximation and leading double logarithm
summation. For the Wilson coefficients $C_i(\mu)$
($i=1,\ldots,10$), we will also use the leading order (LO)
expressions, although the next-to-leading order   calculations
already exist in the literature ~\cite{buras96}. This is the
consistent way to cancel the explicit $\mu$ dependence in the
theoretical formulae.
For the renormalization group evolution of the Wilson coefficients
from higher scale to lower scale, we use the leading logarithmic running equations as given in
Appendix C and D of Ref.~\cite{luy01}.


In the resummation procedures, the $B$ meson is treated as a
heavy-light system. In general, the B meson light-cone matrix
element can be decomposed as ~\cite{grozin}
\begin{eqnarray}
&&\int_0^1\frac{d^4z}{(2\pi)^4}e^{i\bf{k_1}\cdot z}
  \langle 0|\bar{b}_\alpha(0)d_\beta(z)|B(p_B)\rangle \nonumber\\
&=&-\frac{i}{\sqrt{2N_c}}\left\{(\psl_B+m_B)\gamma_5
\left[\phi_B ({\bf k_1})-\frac{\nsl_+- \nsl_-}{\sqrt{2}}
\bar{\phi}_B({\bf k_1})\right]\right\}_{\beta\alpha}, \label{aa1}
\end{eqnarray}
where $n_+=(1,0,{\bf 0_T})$, and $n_-=(0,1,{\bf 0_T})$ are the
unit vectors pointing to the plus and minus directions,
respectively. From the above equation, one can see that there are
two Lorentz structures in the B meson distribution amplitudes.
They obey to the following normalization conditions
 \beq
 \int\frac{d^4 k_1}{(2\pi)^4}\phi_B({\bf k_1})=\frac{f_B}{2\sqrt{2N_c}}, ~~~\int \frac{d^4
k_1}{(2\pi)^4}\bar{\phi}_B({\bf k_1})=0.
 \eeq

In general, one should consider these two Lorentz structures in
calculations of $B$ meson decays. However, it can be argued that
the contribution of $\bar{\phi}_B$ is numerically small
\cite{kurimoto}, thus its contribution can be numerically
neglected safely. Using this approximation, we can reduce one input
parameter in our calculation.  Therefore, we only consider the contribution of Lorentz
structure
\beq
\Phi_B= \frac{1}{\sqrt{2N_c}} (\psl_B +m_B)
\gamma_5 \phi_B ({\bf k_1}). \label{bmeson}
\eeq

The $K$ and $K^*$ mesons are treated as a light-light system.
Based on the SU(3) flavor symmetry, we assume that the wave functions of $K$ and $K^*$ mesons
are the same in structure as the wave functions of $\pi$ and $\rho$, respectively, then
the $K$ meson wave function is defined as \cite{wf,ball}
\beq
\Phi_K(P,x,\zeta)\equiv \frac{1}{\sqrt{2N_c}}\gamma_5  \left\{\psl
\phi_K^{A}(x)+m_0^K \phi_K^{P}(x)+\zeta m_0^K (\vsl \nsl-v\cdot
n)\phi_K^{T}(x)\right\}
\label{eq:pkik}
\eeq
where $P$ and $x$ are the momentum and the momentum fraction of $K$, respectively.
The parameter $\zeta$ is either $+1$ or $-1$ depending on the assignment of the momentum fraction $x$.
While in $B \to K K^*$ decays, $K^*$ meson is longitudinally polarized, only the
longitudinal component $\Phi_{K^*}^L$ of the wave function should be considered ~\cite{kurimoto, bbkt98},
\beq
\Phi_{K^*}^L &=&
\frac{1}{\sqrt{2N_c}} \left\{ \epsl \left [ \psl_{K^*} \phi_{K^*}^T (x) + m_{K^*} \phi_{K^*} (x) \right]
+ m_{K^*} \phi_{K^*}^S (x)\right\}.
\label{eq:phiks}
\eeq
The second term in above equation is the leading twist wave function (twist-2), while the
first and third terms are sub-leading twist (twist-3) wave functions. The transverse part of $\Phi_{K^*}$
can be found for example in Ref.~\cite{zhu05a}.

The explicit expressions of the distribution functions  $\phi_B ({\bf k_1})$, $\phi_K^A(x)$,
$\phi_K^P(x)$, $\phi_K^T(x)$, $\phi_{K^*}(x)$, $\phi_{K^*}^S(x)$ and $\phi_{K^*}^T(x)$ will be given in next
section. The initial conditions of leading twist distribution functions  $\phi_i(x)$,
$i=B, K^*,K$, are of non-perturbative origin, satisfying the normalization condition
\beq
\int_0^1\phi_i(x,b=0)dx=\frac{1}{2\sqrt{6}}{f_i}\;, \label{no}
\eeq
where $f_i$ is the  decay constant of the corresponding meson.

\section{Perturbative Calculations }\label{sec:p-c}

\begin{figure}[tb]
\vspace{-1 cm} \centerline{\epsfxsize=18cm \epsffile{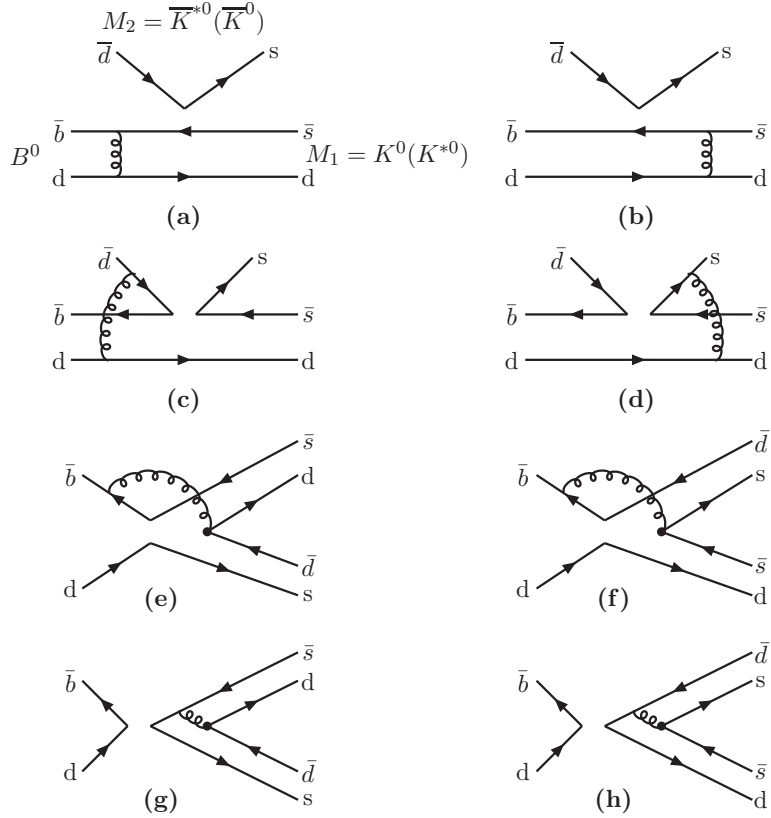}}
\vspace{-14cm} \caption{Typical Feynman diagrams contributing to
$B^0 \to K^0 \overline{K}^{*0}(K^{*0} \overline{K}^0) $ decays. The
diagram (a) and (b) contribute to the form factor $A_0^{B\to K^*}$
or $F_{0,1}^{B\to K}$ for $M_1=K^{*0}$ or $K^0$, respectively. Other
four Feynman diagrams obtained by connecting  the gluon lines to the
d quark line inside the $B^0$ meson for (e) and (f), and  to the
lower s or d quark line for (g) and (h) are omitted.}
\label{fig:fig2}
\end{figure}

\begin{figure}[tb]
\vspace{-1 cm} \centerline{\epsfxsize=18 cm \epsffile{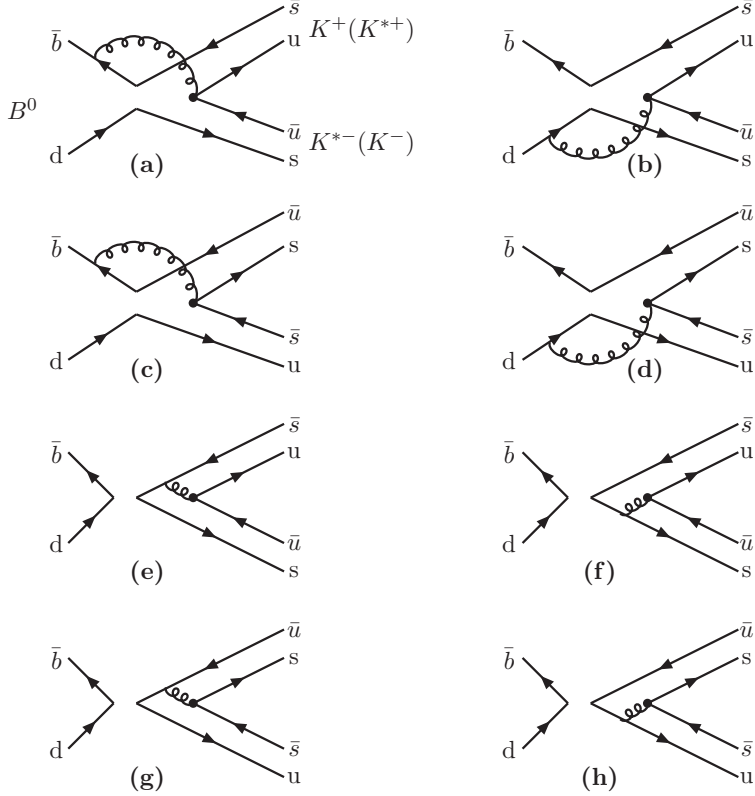}}
\vspace{-13cm} \caption{ Feynman diagrams for $B^0\to K^+
K^{*-}(K^{*+} K^-)$ decays. } \label{fig:fig3}
\end{figure}

\begin{figure}[tb]
\vspace{-1 cm} \centerline{\epsfxsize=18cm \epsffile{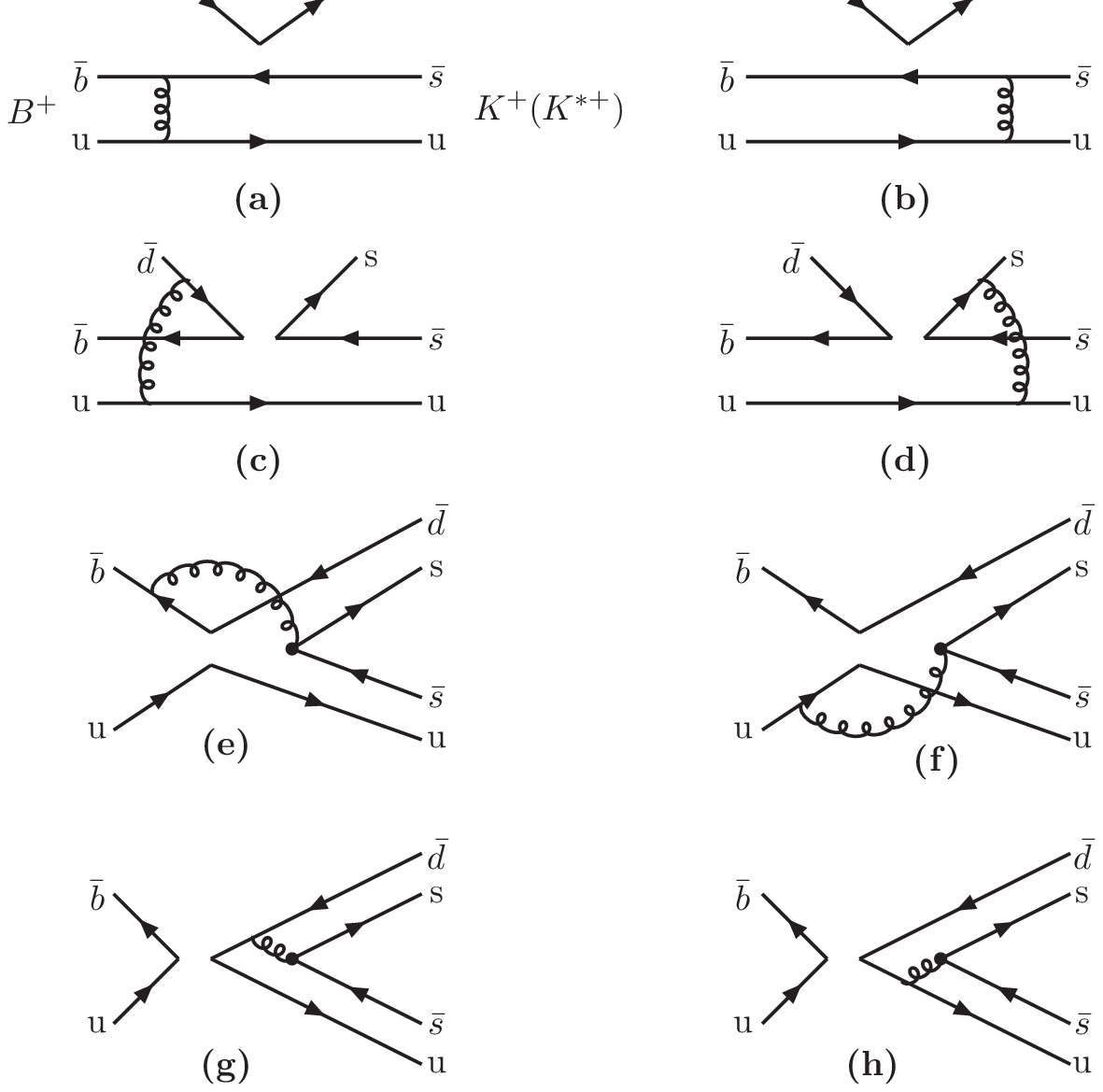}}
\vspace{-13cm} \caption{ Feynman diagrams for $B^+\to K^+ \bar
K^{*0}( K^{*+} \ov K^0)$ decays. }
 \label{fig:fig4}
\end{figure}

For the considered decay modes, the Feynman diagrams are shown in
Figs.~\ref{fig:fig2}-\ref{fig:fig4}. We firstly analyze the
corresponding decay modes topologically: (i) the eight diagrams can
be categorized into emission and annihilation diagrams; (ii) each
category contains four diagrams: two factorizable and two
nonfactorizable. In Fig.~2, for example, Figs.~2(a-d) are emission
diagrams, while Figs.~2(e-h) are annihilation ones topologically;
and Figs. 2(a,b,g,h) are factorizable and Figs.~2(c-f) are
nonfactorizable diagrams.

For $B^0 \to K^0 \overline{K}^{*0}(K^{*0} \overline{K}^0)$ decays,
only the operators $O_{3-10}$ contribute via penguin topology with light quark $q=s$ (diagrams a,b,c,d )
and via the annihilation topology with the light quark $q=d$
(diagram 2(f) and 2(h) ) or $s$ ( diagram 2(e) and 2(g) ).
It is a pure penguin mode with only one kind of CKM elements,
and consequently, there is no CP violation for these decays.

For the $B^0(\bar{B}^0) \to K^{+}K^{*-}(K^{*+}K^-)$ decays ( see Fig.~\ref{fig:fig3}),
the current-current operators $O_{1,2}^{(u)}$ contribute via the
annihilation topology ( Figs.~3(c,d,g,h)), while the operators $O_{3-10}$ contribute via the annihilation
topology with the light quark $q=s$ ( Figs.~3(a,b,e,f))  or  $q =u $ ( Figs.~3(c,d,g,h)).

For the $B^+ \to K^+ \bar{K}^{*0}(K^{*+} \bar{K}^0)$ decays ( see
Fig.~\ref{fig:fig4}), the current-current operators $O_{1,2}^{(u)}$
contribute via the annihilation topology ( Figs.~4(e-h)), while the
penguin operators $O_{3-10}$ contribute via the penguin topology
with the light quark $q=s$ (Figs.~4(a-d)) or via the annihilation
topology with $q=u$ ( Figs.~4(e-h)).

In the analytic calculations, the operators with $(V-A)(V-A)$
structure work directly, while the operators with $(V-A)(V+A)$
structure will work in two different ways:
\begin{itemize}
\item
In some decay channels, some of these operators contribute directly
to the decay amplitude in a factorizable way.

\item
In some other cases, we need to do Fierz transformation for these operators to get right
flavor and color structure for factorization to work. In this case, we get $(S+P)(S-P)$ operators
from $(V-A)(V+A)$ ones.
\end{itemize}

\subsection{$B^0 \to K^0 \overline{K}^{*0}(K^{*0} \overline{K}^0)$ decay}

For the sake of the reader, we take the $B^0 \to K^0
\overline{K}^{*0}(K^{*0} \overline{K}^0)$ decay channel as an
example to show the ways to derive the decay amplitude from
individual diagram. As shown explicitly in Fig.~2(a), the meson
$M_1$ which picks up the spectator quark can be $K^0$ or $K^{*0}$,
the emitted meson $M_2$ should be $\overline{K}^{*0}$ or
$\overline{K}^{0}$ at the same time. The $B^0$ meson therefore can
decay into the final state $f=K^0 \overline{K}^{*0}$ and $\bar{f}=
K^{*0} \overline{K}^{0}$ simultaneously. The $\overline{B}^0$ meson,
on the other hand, also decay into the same final state $f=K^0
\overline{K}^{*0}$ and $\bar{f}= K^{*0} \overline{K}^{0}$
simultaneously.

Now we consider the usual factorizable diagram 2(a) and 2(b) for the
case of $M_1=K^{*0}$. The $(V-A)(V-A)$ operators $O_{3,4}$ and
$O_{9,10}$ contribute through diagram 2(a) and 2(b), the sum of
their contributions is given as \beq F_{eK^*}&=& 4 \sqrt{2} G_F \pi
C_F f_K m_B^4 \int_0^1 d x_{1} dx_{3}\, \int_{0}^{\infty} b_1 db_1
b_3 db_3\, \phi_B(x_1,b_1)  \non & & \cdot \left\{ \left [(1+x_3)
\phi_{K^*} (x_3, b_3) +(1-2x_3)r_{K^*} (\phi_{K^*}^s (x_3,
b_3)+\phi_{K^*}^t (x_3,b_3)) \right] \right.  \non
 && \left. \cdot \alpha_s(t_e^1) h_e(x_1,x_3,b_1,b_3)\exp[-S_{a}(t_e^1)]
 \right.   \non
&& \left. +2 r_{K^*} \phi_{K^*}^s (x_3, b_3) \alpha_s(t_e^2)
h_e(x_3,x_1,b_3,b_1)\exp[-S_{a}(t_e^2)] \right\} \;,
\label{eq:feks}
\eeq
where  $C_F=4/3$ is a color factor. The functions $h_e^i$, the
scales $t_e^i$ and the Sudakov factors $S_{a}(t_e^1)$ and $S_{a}(t_e^2)$ will be given
explicitly in the Appendix. In Eq.~(\ref{eq:feks}), we do not include the Wilson
coefficients of the corresponding operators, which are process dependent.
They will be shown later in this section for different decay channels.

The form factor of $B$ to $K^*$ transition, $A_0^{B\to K^*}(0)$, can
also be extracted from $F_{eK^*}$ in Eq.~(\ref{eq:feks}), that is
\beq
A_0^{B\to K^*}(q^2=0)=\frac{\sqrt{2}~F_{eK^*}} {G_F f_K m_B^2}.
\label{eq:f0bks}
\eeq

The operators $O_{5-8}$ have a structure of $(V-A)(V+A)$. Some of these operators contribute
to the decay amplitude in a factorizable way. Since only
the axial-vector part of $(V+A)$ current contribute to the
pseudo-scaler meson production,
 \beq
\langle K^* |V-A|B\rangle \langle K |V+A | 0 \rangle = -\langle K^*
|V-A |B  \rangle \langle K |V-A|0 \rangle .
 \eeq
The contribution of these operators is opposite in sign with $F_{eK^*}$ in Eq.~(\ref{eq:feks}):
\beq
F_{eK^*}^{P1} = -F_{eK^*}.\label{eq:feksp1}
\eeq

In some other cases, one needs to do Fierz transformation  for these operators first and then
get right color structure for factorization to work. In this case, one gets $(S-P)(S+P)$ operators
from $(V-A)(V+A)$ ones. For these $(S-P)(S+P)$ operators, Figs.~ 2(a) and 2(b) gives
\beq
F_{eK^*}^{P_2}&=& 8 \sqrt{2} G_F \pi C_F f_K r_K m_B^4 \int_{0}^{1}d x_{1}d
x_{3}\,\int_{0}^{\infty} b_1d b_1 b_3d b_3\, \phi_B(x_1,b_1) \non &
& \cdot
 \left\{ \left[ \phi_{K^*} (x_3, b_3)+r_{K^*} ((x_3+2) \phi_{K^*}^s (x_3, b_3)- x_3 \phi_{K^*}^t (x_3,
 b_3))\right]\right.
\non
 & &\left. \cdot \alpha_s (t_e^1)  h_e
(x_1,x_3,b_1,b_3)\exp[-S_{a}(t_e^1)]
  \right.  \non
& &\left.  + \left(x_1 \phi_{K^*}(x_3,b_3)+ 2 r_{K^*} \phi_{K^*}^s
(x_3, b_3)\right) \alpha_s (t_e^2)
 h_e(x_3,x_1,b_3,b_1)\exp[-S_{a}(t_e^2)] \right\} \;.
\label{eq:feksp2}
\eeq

For the non-factorizable diagram 2(c) and 2(d), all three meson wave
functions are involved. The integration of $b_3$ can be performed
using $\delta$ function $\delta(b_3-b_1)$, leaving only integration
of $b_1$ and $b_2$. $M_{eK^*}$ denotes the contribution from the
operators of type $(V-A)(V-A)$, and $M_{eK^*}^{P_1}$ is the
contribution from the operators of type $(V-A)(V+A)$:
\beq
M_{eK^*}&=& \frac{16} {\sqrt{3}} G_F \pi C_F m_B^4 \int_{0}^{1}d
x_{1}d x_{2}\,d x_{3}\,\int_{0}^{\infty} b_1d b_1 b_2d b_2\,
\phi_B(x_1,b_1) \phi_K^A(x_2,b_2) \non
 & &\cdot
\left
\{-\left[-x_2\phi_{K^*}(x_3,b_1)+r_{K^*}x_3\left(\phi_{K^*}^s(x_3,b_1)-\phi_{K^*}^t(x_3,b_1)\right)\right]
\right.
 \non
 & &\left.\cdot
\alpha_s(t_f)
 h_f^1(x_1,x_2,x_3,b_1,b_2)\exp[-S_{c}(t_f^1)]
 \right.
 \non
 & &\left.+
 \left[(x_2-x_3-1)\phi_{K^*}(x_3,b_1)+r_{K^*}x_3\left(\phi_{K^*}^s(x_3,b_1)+\phi_{K^*}^t(x_3,b_1)\right)
\right] \right.  \non
 & &\left.\cdot
 \alpha_s(t_f)
 h_f^2(x_1,x_2,x_3,b_1,b_2)\exp[-S_{c}(t_f^2)]\right \} \; ,
\label{eq:meks}
 \eeq
\beq
M_{eK^*}^{P_1}&=& \frac{16} {\sqrt{3}} G_F \pi C_F m_B^4
\int_{0}^{1}d x_{1}d x_{2}\,d x_{3}\,\int_{0}^{\infty} b_1d b_1 b_2d
b_2\, \phi_B(x_1,b_1) r_K \non & &\cdot \left
\{\left[(x_1-x_2)\left(\phi_K^P(x_2,b_2)-\phi_K^T(x_2,b_2)\right)\phi_{K^*}(x_3,b_1)+r_{K^*}
\left(x_1\left(\phi_K^P(x_2,b_2)
 \right.\right.\right.\right.
 \non
& &\left.\left.\left.\left.
-\phi_K^T(x_2,b_2)\right)\left(\phi_{K^*}^s(x_3,b_1)-\phi_{K^*}^t(x_3,b_1)\right)
-x_2\left(\phi_K^P(x_2,b_2)-\phi_K^T(x_2,b_2)\right)
\right.\right.\right.
 \non
& &\left.\left.\left.\cdot
\left(\phi_{K^*}^s(x_3,b_1)-\phi_{K^*}^t(x_3,b_1)\right)-x_3\left(\phi_K^P(x_2,b_2)
+\phi_K^T(x_2,b_2)\right)\left(\phi_{K^*}^s(x_3,b_1)
 \right.\right.\right.\right.  \non
& &\left.\left.\left.\left.
+\phi_{K^*}^t(x_3,b_1)\right)\right)\right]\alpha_s(t_f)
 h_f^1(x_1,x_2,x_3,b_1,b_2)\exp[-S_{c}(t_f^1)]
\right.  \non
& &\left.
 -\left[(x_1+x_2-1)\left(\phi_K^P(x_2,b_2)-\phi_K^T(x_2,b_2)\right)
 \phi_{K^*}(x_3,b_1)
\right.\right.
 \non
& &\left.\left.
+r_{K^*}\left(x_1\left(\phi_K^P(x_2,b_2)-\phi_K^T(x_2,b_2)\right)
\left(\phi_{K^*}^s(x_3,b_1)-\phi_{K^*}^t(x_3,b_1)\right)
\right.\right.\right.
 \non
& &\left.\left.\left.
-(1-x_2)\left(\phi_K^P(x_2,b_2)-\phi_K^T(x_2,b_2)\right)\left(\phi_{K^*}^s(x_3,b_1)-\phi_{K^*}^t(x_3,b_1)\right)
\right.\right.\right.
 \non
& &\left.\left.\left.
-x_3\left(\phi_K^P(x_2,b_2)+\phi_K^T(x_2,b_2)\right)\left(\phi_{K^*}^s(x_3,b_1)
+\phi_{K^*}^t(x_3,b_1)\right)\right)\right]
 \right.  \non & &\left.
 \alpha_s(t_f)
 h_f^2(x_1,x_2,x_3,b_1,b_2)\exp[-S_{c}(t_f^2)]\right \} \; .
\label{eq:meksp1}
 \eeq

For the non-factorizable annihilation diagram 2(e), we have three kinds of
contributions: $M_{a{K^{*}}}$ for $(V-A)(V-A)$ operators,
$M_{aK^{*}}^{P_1}$ for $(V-A)(V+A)$ operators and
$M_{a{K^*}}^{P_2}$ for $(S-P)(S+P)$ operators.
\beq
M_{aK^*}&=& \frac{16} {\sqrt{3}} G_F \pi C_F m_B^4 \int_{0}^{1}d
x_{1}d x_{2}\,d x_{3}\,\int_{0}^{\infty} b_1d b_1 b_2d b_2\,
\phi_B(x_1,b_1) \non && \cdot \left \{ -\left[x_2
\phi_{K^*}(x_3,b_2) \phi_K^A(x_2,b_2) + r_{K^*} r_K
\left(\phi_K^P(x_2,b_2)\left(\left(x_2+x_3+2.\right)
\right.\right.\right.\right.   \non
& &\left.\left.\left.\left.
 \cdot\phi_{K^*}^s(x_3,b_2)+\left(x_2-x_3\right)\phi_{K^*}^t(x_3, b_2)\right)+\phi_K^T(x_2,b_2)\left(-x_3\left
(\phi_{K^*}^s(x_3,b_2) \right.\right.\right.\right.\right.
  \non
& &\left.\left.\left.\left.\left.
-\phi_{K^*}^t(x_3,b_2)\right)-2\phi_{K^*}^t(x_3,b_2)+x_2
\left(\phi_{K^*}^s(x_3,b_2)+\phi_{K^*}^t(x_3,b_2)
 \right)\right)\right)\right ]\right.
  \non
& & \left.
\cdot\alpha_s(t_f^3)h_f^3(x_1,x_2,x_3,b_1,b_2)\exp[-S_{c}(t_f^3)]~
\right.
  \non
& & \left. +~\left[ x_3 \phi_{K^*}(x_3, b_2) \phi_K^A(x_2,b_2) -
r_{K^*} r_K
\left(-x_2\left(\phi_K^P(x_2,b_2)-\phi_K^T(x_2,b_2)\right)
\right.\right.\right.
 \non
& &\left.\left.\left.
\cdot\left(\phi_{K^*}^s-\phi_{K^*}^t\right)-x_3\left(\phi_K^P(x_2,b_2)+\phi_K^T(x_2,b_2)
\right)\left(\phi_{K^*}^s(x_3, b_2)+\phi_{K^*}^t(x_3,
b_2)\right)\right) \right]\right.
 \non
&&\left.\cdot  \alpha_s(t_f^4)
h_f^4(x_1,x_2,x_3,b_1,b_2)\exp[-S_{c}(t_f^4)]~\right\}
 \;,
\eeq
where $r_K= m_0^K/m_B $ with $m_0^K=m_K^2/(m_s+m_d)$.
\beq
M_{a{K^*}}^{P_1}&=& \frac{16} {\sqrt{3}} G_F \pi C_F
m_B^4 \int_{0}^{1}d x_{1}d x_{2}\,d x_{3}\,\int_{0}^{\infty} b_1d
b_1 b_2d b_2\, \phi_B(x_1,b_1)
 \non
& &\cdot
\left\{\left[-r_{K^*}\left(x_3-2\right)\left(\phi_{K^*}^s(x_3,b_2)
+\phi_{K^*}^t(x_3,b_2)\right) \right.\right.
 \non
 & &\left.\left.+r_K\left(x_2-2\right)\phi_{K^*}(x_3,b_2)
\left (\phi_K^P(x_2,b_2)+\phi_K^T(x_2,b_2) \right)\right] \right.
 \non
 & &\left.\cdot
\alpha_s(t_f^3)h_f^3(x_1,x_2,x_3,b_1,b_2)\exp[-S_{c}(t_f^3)] \right.
\non & &\left.+\left[-x_2 r_K \phi_{K^*}(x_3, b_2) \left
(\phi_K^P(x_2,b_2)+\phi_K^T(x_2,b_2)\right)
\right.\right.
 \non
 & &\left.\left. +x_3r_{K^*} \phi_K^A(x_2,b_2)\left(\phi_{K^*}^s(x_3,b_2)+\phi_{K^*}^t(x_3,b_2)
\right) \right]\right.
 \non
 & &\left. \cdot \alpha_s(t_f^4)h_f^4(x_1,x_2,x_3,b_1,b_2)
\exp[-S_{c}(t_f^4)]\right\} \;.
 \eeq

 \beq
 M_{a{K^*}}^{P_2}&=& \frac{16} {\sqrt{3}} G_F \pi C_F
m_B^4 \int_{0}^{1}d x_{1}d x_{2}\,d x_{3}\,\int_{0}^{\infty} b_1d
b_1 b_2d b_2\, \phi_B(x_1,b_1)
 \non
& &\cdot
\left\{\left[x_3\phi_{K^*}(x_3,b_2)\phi_K^A(x_2,b_2)+r_{K^*}r_K\left(\left(
\left(x_2+x_3+2\right)\phi_{K^*}^s(x_3,b_2)
\right.\right.\right.\right.
 \non
 & &
\left.\left.\left.\left.-\left(x_2-x_3\right)
 \phi_{K^*}^t(x_3,b_2)\right)\phi_K^P(x_2,b_2)+\left(x_3\left(\phi_{K^*}^s(x_3,b_2)
+\phi_{K^*}^t(x_3,b_2)\right)
 \right.\right.\right.\right.
 \non
 & &
\left.\left.\left.\left. +x_2\left(\phi_{K^*}^t(x_3,b_2)
-\phi_{K^*}^s(x_3,b_2)\right)-2\phi_{K^*}^t(x_3,b_2)\right)\phi_K^T(x_2,b_2)\right)\right]
\right. \non
 & &
\cdot\left.
\alpha_s(t_f^3)h_f^3(x_1,x_2,x_3,b_1,b_2)\exp[-S_{c}(t_f^3)] \right.
\non
&&\left.+\left[-x_2\phi_{K^*}(x_3,b_2)\phi_K^A(x_2,b_2)+r_{K^*}r_K
\left(-x_2\left(\phi_K^P(x_2,b_2)+\phi_K^T(x_2,b_2)\right)
 \right.\right.\right.\non
 & &
 \cdot\left.\left.\left.
 \left(\phi_{K^*}^s(x_3,b_2)+\phi_{K^*}^t(x_3,b_2)\right)-x_3\left(\phi_K^P(x_2,b_2)-\phi_K^T(x_2,b_2)\right)
\right.\right.\right.\non
 & &
 \cdot\left.\left.\left.
\left(\phi_{K^*}^s(x_3,b_2)-\phi_{K^*}^t(x_3,b_2)\right)\right)\right]
\alpha_s(t_f^4)h_f^4(x_1,x_2,x_3,b_1,b_2)
\exp[-S_{c}(t_f^4)]\right\} \;.
 \eeq

The factorizable annihilation diagram 2(g) involves only $K^*$ and
$K$ wave functions. The decay amplitude $F_{aK^*}$, $F_{aK^*}^{P1}$ and $F_{aK^*}^{P2}$
represent the contributions from $(V-A)(V-A)$ operators, $(V-A)(V+A)$ operators and
$(S-P)(S+P)$ operators, respectively.
\beq
F_{aK^*}&=& -4 \sqrt{2} \pi G_F C_F f_B m_B^4 \int_{0}^{1}d
x_{2}\,d x_{3}\,\int_{0}^{\infty} b_2d b_2b_3d b_3 \,
 \non
& &\cdot \left\{ \left[ x_3 \phi_{K^*}(x_3,b_3) \phi_K^A(x_2,b_2) +
2 r_{K^*} r_K \phi_K^P(x_2,b_2) \left((1+x_3)\phi_{K^*}^s(x_3,
b_3)\right.\right.\right. \non
 &&\left.\left.\left.-(1-x_3) \phi_{K^*}^t(x_3,b_2) \right)
\right]\alpha_s(t_e^3)
h_a(x_2,x_3,b_2,b_3)\exp[-S_{d}(t_e^3)]\right. \non
 &&\left. -\left[x_2 \phi_{K^*}(x_3,b_3) \phi_K^A(x_2,b_2)+2 r_{K^*}
r_K
\phi_{K^*}^s(x_3,b_3)\left((1+x_2)\phi_K^P(x_2,b_2)\right.\right.\right.
 \non
& &\left.\left.\left.-(1-x_2)\phi_K^T(x_2,b_2) \right) \right]
 \alpha_s(t_e^4)
 h_a(x_3,x_2,b_3,b_2)\exp[-S_{d}(t_e^4)]\right \}\;,
 \label{eq:faks}\\
F_{a{K^*}}^{P_1}&=&-F_{a{K^*}},
\label{eq:faksp1}
\eeq
\beq
F_{a{K^*}}^{P_2} &=& -8 \sqrt{2} G_F \pi C_F  m_B^4 f_B \int_{0}^{1}d x_{2}\,d
x_{3}\,\int_{0}^{\infty} b_2d b_2b_3d b_3 \,
 \non
& &\cdot \left\{ \left[2 r_K \phi_{K^*}(x_3, b_3) \phi_K^P(x_2,b_2)
+ x_3 r_{K^*} \left(\phi_{K^*}^s(x_3, b_3)- \phi_{K^*}^t(x_3,b_2)
\right) \phi_K^A(x_2,b_2) \right]\right.
 \non
&&\left.\cdot \alpha_s(t_e^3)
h_a(x_2,x_3,b_2,b_3)\exp[-S_{d}(t_e^3)]\right.
 \non
 &&\left.+\left[2 r_{K^*}
\phi_{K^*}^s(x_3,b_3) \phi_K^A(x_2,b_2)+ x_2 r_K
 \left
(\phi_K^P(x_2,b_2)-\phi_K^T(x_2,b_2)\right)\phi_{K^*}(x_3,b_3)
\right]\right.  \non
 &&\left. \cdot
 \alpha_s(t_e^4)
 h_a(x_3,x_2,b_3,b_2)\exp[-S_{d}(t_e^4)]\right \}\;. \label{eq:faksp2}
\eeq

In the above equations, we have assumed that $x_1 <<x_2,x_3$. Since
the light quark momentum fraction $x_1$ in $B$ meson is peaked at
the small $x_1$ region, while quark momentum fraction $x_2$ of $K$
is peaked around $0.5$, this is not a bad approximation. The
numerical results also show that this approximation makes very
little difference in the final result. After using this
approximation, all the diagrams are functions of $k_1^-= x_1 m_B/\sqrt{2}$ of B meson only,
independent of the variable of $k_1^+$.

For the Feynman diagram 2(f) and 2(h), the corresponding decay amplitude is the same in structure
as those for 2(e) and 2(g). We get the decay amplitude easily by making two replacements of
$x_2 \to 1-x_2$ and $x_3 \to 1-x_3$ in the relevant distribution amplitudes.

For the case of $M_1=K^0$ and $M_2=\overline{K}^{*0}$,
by following the same procedure, one can find all decay amplitudes:
$F_{eK}$, $F_{eK}^{P_1}$ and $F_{eK}^{P_2}$, $M_{eK}$,
$M_{eK}^{P_1}$, $M_{aK}$,  $M_{aK}^{P_1}$ and $M_{aK}^{P_2}$,
$F_{aK}$, $F_{aK}^{P_1}$ and $F_{aK}^{P_2}$.
The explicit  expressions of these decay amplitudes will be given in Appendix A.

\subsection{Total Decay amplitudes}

Based the isospin symmetry and the analytical results obtained in last subsection,
one can derive out all the decay amplitudes for
$B^0\to K^+ K^{*-} (K^{*+} K^-)$ and $B^+\to K^+ \bar K^{*0} ( K^{*+} \ov K^0)$ decays.

Combining all contributions, the total decay amplitude for all
considered decay modes can be written as: {\footnotesize \beq
 {\cal M} (B^0\to K^0 \bar K^{*0}) &=&
 -\xi_t \left \{ F_{eK}\left( \frac{C_3}{3}
+C_4-\frac{C_9}{6}-\frac{C_{10}}{2}\right)
+M_{eK}\left(C_3-\frac{C_9}{2}\right)+M_{eK}^{P_1}\left(C_5-\frac{C_7}{2}\right)
\right.
 \non
&&\hspace{-30mm}\left.+M_{aK}\left(C_3+C_4-\frac{C_9}{2}-\frac{C_{10}}{2}\right)
+M_{aK}^{P_1}\left(C_5-\frac{C_7}{2}\right)
 +M_{aK}^{P_2}\left(C_6-\frac{C_8}{2}\right)+M_{aK^*}\left(C_4-\frac{C_{10}}{2}\right)
\right.
 \non
 &&\hspace{-30mm}\left.+F_{aK}\left(\frac{4}{3}C_3+\frac{4}{3}C_4-C_5-\frac{C_6}{3}+\frac{C_7}{2}+\frac{C_8}{6}
 -\frac{2}{3}C_9-\frac{2}{3}C_{10}\right)+M_{aK^*}^{P_2}\left(C_6-\frac{C_8}{2}\right)
 \right.
 \non
&&\hspace{-30mm}\left.+F_{aK^*}\left(C_3+\frac{C_4}{3}-C_5-\frac{C_6}{3}+\frac{C_7}{2}+\frac{C_8}{6}
 -\frac{C_9}{2}-\frac{C_{10}}{6}\right)+F_{aK}^{P_2}\left(\frac{C_5}{3}+C_6-\frac{C_7}{6}-\frac{C_8}{2}\right)\right \}  \label{eq:m1a} \eeq

 \beq
{\cal M}(B^0\to K^{*0} \overline{K}^0) &=&  -\xi_t \left \{
F_{eK^*}\left( \frac{C_3}{3}
+C_4-\frac{C_9}{6}-\frac{C_{10}}{2}\right)+F_{eK^*}^{P_2}\left(\frac{C_5}{3}+C_6-\frac{C_7}{6}-\frac{C_8}{2}\right)
 \right. \non
& &\hspace{-30mm}\left.
+M_{eK^*}\left(C_3-\frac{C_9}{2}\right)+M_{eK^*}^{P_1}\left(C_5-\frac{C_7}{2}\right)
+M_{aK^*}\left(C_3+C_4-\frac{C_9}{2}-\frac{C_{10}}{2}\right) \right.
 \non
 &&\hspace{-30mm}\left.+M_{aK^*}^{P_1}\left(C_5-\frac{C_7}{2}\right)
 +M_{aK^*}^{P_2}\left(C_6-\frac{C_8}{2}\right)+F_{aK^*}^{P_2}\left(\frac{C_5}{3}+C_6-\frac{C_7}{6}-\frac{C_8}{2}\right)
\right.
 \non
&&\hspace{-30mm}\left.+F_{aK}\left(C_3+\frac{C_4}{3}-C_5-\frac{C_6}{3}+\frac{C_7}{2}+\frac{C_8}{6}
 -\frac{C_9}{2}-\frac{C_{10}}{6}\right)+M_{aK}\left(C_4-\frac{C_{10}}{2}\right)\right. \non
&&\hspace{-30mm}\left.+F_{aK^*}\left(\frac{4}{3}C_3+\frac{4}{3}C_4-C_5-\frac{C_6}{3}+\frac{C_7}{2}+\frac{C_8}{6}
 -\frac{2}{3}C_9-\frac{2}{3}C_{10}\right)
+M_{aK}^{P_2}\left(C_6-\frac{C_8}{2}\right) \right \} \label{eq:m1b}
\eeq

 \beq
{\cal M}(B^0\to K^+ K^{*-}) &=&\displaystyle \xi_u \left
[M_{aK}C_2+F_{aK}\left(C_1+\frac{C_2}{3}\right)\right ]
 -\xi_t \left \{
 M_{aK}\left(C_4+C_{10}\right)+M_{aK^*}^{P_2}\left(C_6-\frac{C_8}{2}\right)\right. \non
 & &\hspace{-30mm}\left.+F_{aK}\left(C_3+\frac{C_4}{3}-C_5-\frac{C_6}{3}-C_7-\frac{C_8}{3}+C_9+\frac{C_{10}}{3}\right)
 +M_{aK}^{P_2}\left(C_6+C_{8}\right)
\right.  \non
 &&\hspace{-30mm}\left.+F_{aK^*}\left(C_3+\frac{C_4}{3}-C_5-\frac{C_6}{3}+\frac{C_7}{2}+\frac{C_8}{6}
 -\frac{1}{2}C_9-\frac{C_{10}}{6}\right)+ M_{aK^*}\left(C_4-\frac{C_{10}}{2}\right)\right\}\label{eq:m2a} \eeq
\beq {\cal M}(B^0 \to K^{*+} K^- ) &=& \xi_u \left
[M_{aK^*}C_2+F_{aK^*}\left(C_1+\frac{C_2}{3}\right)\right] \non
 & &\hspace{-30mm}
 -\xi_t \left \{ F_{aK}\left(C_3+\frac{C_4}{3}-C_5-\frac{C_6}{3}+\frac{C_7}{2}+\frac{C_8}{6}
-\frac{C_9}{2}-\frac{C_{10}}{6}\right) \right.  \non
 &&\hspace{-30mm}\left.+F_{aK^*}\left(C_3+\frac{C_4}{3}-C_5-\frac{C_6}{3}-C_7-\frac{C_8}{3}+C_9+\frac{C_{10}}{3}\right)
  + M_{aK^*}\left(C_4+C_{10}\right)\right.  \non
& &\hspace{-30mm}\left. +
M_{aK}\left(C_4-\frac{C_{10}}{2}\right)+M_{aK}^{P_2}\left(C_6-\frac{C_{8}}{2}\right)
+M_{aK^*}^{P_2}\left(C_6+C_8\right)
 \right \}
\label{eq:m2b}
 \eeq

 \beq
 {\cal M}(B^+ \to K^+ \ov K^{*0}) &=& \xi_u \left(M_{aK}C_1+F_{aK}\left(\frac{C_1}{3}+C_2\right)\right)
  -\xi_t \left \{ F_{eK}\left(\frac{C_3}{3}+C_4-\frac{C_9}{6}-\frac{C_{10}}{2}\right)
\right.  \non
 &&\hspace{-30mm}\left.+F_{aK}^{P_2}\left(\frac{C_5}{3}+C_6+\frac{C_7}{3}+C_8\right)
+M_{eK}\left(C_3-\frac{C_9}{2}\right)+M_{eK}^{P_1}\left(C_5-\frac{C_7}{2}\right)
 \right.  \non
&&\hspace{-30mm}\left.+M_{aK}\left(C_3+C_9\right)+M_{aK}^{P_1}\left(C_5+C_7\right)
+F_{aK}\left(\frac{C_3}{3}+C_4+\frac{C_9}{3}+C_{10}\right)
 \right \}\label{eq:m3a}
\eeq

\beq {\cal M}(B^+ \to K^{*+} \ov K^0) &=& \xi_u
\left(M_{aK^*}C_1+F_{aK^*}\left(\frac{C_1}{3}+C_2\right)\right)
  -\xi_t \left \{ F_{eK^*}\left(\frac{C_3}{3}+C_4-\frac{C_9}{6}-\frac{C_{10}}{2}\right)
  \right. \non
 &&\hspace{-30mm}\left.+F_{eK^*}^{P_2}\left(\frac{C_5}{3}+C_6-\frac{C_7}{6}-\frac{C_8}{2}\right)
+F_{aK^*}^{P_2}\left(\frac{C_5}{3}+C_6+\frac{C_7}{3}+C_8\right)+M_{eK^*}\left(C_3-\frac{C_9}{2}\right)
 \right.  \non
&&\hspace{-30mm}\left.+M_{eK^*}^{P_1}\left(C_5-\frac{C_7}{2}\right)+M_{aK^*}\left(C_3+C_9\right)+M_{aK^*}^{P_1}\left(C_5+C_7\right)
+F_{aK^*}\left(\frac{C_3}{3}+C_4+\frac{C_9}{3}+C_{10}\right)
 \right \}
\label{eq:m3b} \eeq

}
\par\noindent
where $\xi_u = V_{ub}^*V_{ud}, \xi_t = V_{tb}^*V_{td}$. The exact
expressions of individual transition amplitudes not given explicitly
in this section, such as  $F_{aK}$ and $M_{aK}$, etc., are collected in the Appendix A.

The decay amplitudes for those charge-conjugated decay channels can be obtained from
the results as given in Eqs.(\ref{eq:m1a}) - (\ref{eq:m3b}) by simple replacements of
$\xi_u \to \xi_u^*$ and $\xi_t \to  \xi_t^*$.

Analogous to Eq.~(\ref{eq:f0bks}), the form factor $F_{0,1}^{B\to K}(q^2=0)$
can also be extracted from $F_{eK}$ via the following relation
\beq
F_{0,1}^{B\to K}(q^2=0)=\frac{\sqrt{2}~F_{eK}} {G_F f_{K^*} m_B^2}.
\label{eq:f01}
\eeq

\section{Numerical results and Discussions}\label{sec:n-d}

\subsection{Input parameters and wave functions}

Before we calculate the branching ratios and CP violating asymmetries for the B decays under study,
we firstly present the input parameters to be used in the numerical calculations.
\beq
\Lambda_{\overline{\mathrm{MS}}}^{(f=4)} &=& 0.25 {\rm GeV}, \quad
f_B = 0.19 {\rm GeV},\quad  m_0^K= 1.7 {\rm GeV},\non
f_{K^*} &=& 0.217 {\rm GeV},\quad f_{K^*}^T = f_K = 0.16  {\rm GeV},
\quad m_K=0.497 {\rm GeV},   \non
m_{K^*}&=&0.89{\rm GeV}, \quad  M_B = 5.2792 {\rm GeV}, \quad M_W = 80.41{\rm GeV}.
 \label{para}
\eeq
The central values of the CKM matrix elements to be used in
numerical calculations are
\beq
|V_{ud}|&=&0.9745,
\quad |V_{ub}|=0.0036,\non |V_{tb}|&=&0.9990, \quad |V_{td}|=0.0075.
\eeq

For the $B$ meson wave function, we adopt the model\cite{chen00a,luy01,kurimoto}
\beq
\phi_B(x,b) &=&  N_B x^2(1-x)^2 \mathrm{exp} \left
 [ -\frac{M_B^2\ x^2}{2 \omega_{b}^2} -\frac{1}{2} (\omega_{b} b)^2\right],
 \label{phib}
\eeq
where the shape parameter $\omega_{b}=0.4\pm 0.04$ GeV has been constrained in other decay modes.
The normalization constant $N_B=91.745$ is related to $f_B=0.19$GeV and $\omega_{b}=0.4$.

The $K^*$ meson distribution amplitude up to twist-3
are given by \cite{bbkt98} with QCD sum rules.
\beq
\phi_{K^*}(x) &=& \frac{3}{\sqrt{6} }
 f_{K^*}  x (1-x)  \left[1+ 0.57(1-2x)+0.07C_2^{3/2} (1-2x) \right],\\
\phi_{K^*}^t(x) &=&  \frac{f_{K^*}^T }{2\sqrt{6} }
  \left\{  0.3 (1-2 x)\left(3(1-2x)^2+10(1-2x)-1\right)  \right.  \non
 &&~~\left. +1.68C_4^{1/2}(1-2x)+0.06 (1-2 x)^2\left(5 (1-2 x)^2-3\right)
 \right. \non
 &&~~\left.+0.36\left[1-2 (1-2 x)-2 (1-2 x)\ln(1-x)\right]\right\},\\
\phi_{K^*}^s(x) &=&  \frac{f_{K^*}^T}{2\sqrt{6} }
  \left\{3(1-2x)\left[1+0.2(1-2x)+ 0.6 (10 x^2 -10 x +1) \right]
  \right.  \non
 &&~~\left. -0.12x(1-x) + 0.36\left[1-6x-2\ln(1-x)\right]\right\},
\eeq
where the Gegenbauer polynomials are defined by
 \beq
 C_2^{3/2} (t) = \frac{3}{2} \left (5t^2-1 \right ),
 \quad C_4^{1/2} (t) = \frac{1}{8} \left (35t^4-30t^2+3 \right ).
\eeq

For $K$ meson, we use $\phi_K^A$ of twist-2 wave function and  $\phi_K^P$ and $\phi_K^T$
of the twist-3 wavefunctions from  \cite{bbkt98,ball}
 \beq
 \phi_K^A(x)&=&\frac{3}{\sqrt{6}}f_Kx(1-x) \left [ 1+0.51(1-2x)+0.3\left(5(1-2x)^2-1\right)\right ],\\
\phi^P_K(x)&=&\frac{f_K}{2\sqrt{6}} \left[1+0.12(3(1-2x)^2-1)
\right.
  \nonumber\\
& &\left. -0.12\left(3-30(1-2x)^2+35(1-2x)^4\right)/8\right],\\
\phi^T_K(x) &=&\frac{f_K}{2\sqrt{6}}(1-2x) \left
[1+0.35\left(10x^2-10x+1\right)\right] .
 \eeq

Based on the definition of the form factor $A_0^{B \to K^*}$ and
$F_{0,1}^{B \to K}$ as given in Eqs.~(\ref{eq:f0bks}) and (\ref{eq:f01}),
we find the numerical values of the corresponding
form factors at zero momentum transfer.
\beq
A_0^{B\to K^*}(q^2=0)&=& 0.46^{+0.07}_{-0.06}(\omega_b),\non
F_{0,1}^{B \to K}(q^2=0)&=& 0.35^{+0.06}_{-0.04}(\omega_b). \label{eq:aff0}
\eeq
where the errors are induced by the change of $\omega_b$ for $\omega_b=0.40 \pm 0.04$ GeV.
These results are close to the light-cone QCD sum rule predictions \cite{ball05}
\beq
A_0^{B\to K^*}(q^2=0)&=& 0.374\pm 0.034,\non
F_{0,1}^{B \to K}(q^2=0)&=& 0.331\pm 0.041
\eeq

\subsection{Branching ratios}

In order to calculate the branching ratios and CP asymmetries in a more clear way, we rewrite
the decay amplitudes as given in Eqs.(\ref{eq:m1a})-(\ref{eq:m3b}) in a new form
 \beq
{\cal M} &=& V_{ub}^*V_{ud} T -V_{tb}^* V_{td} P= V_{ub}^*V_{ud} T
\left [ 1 + z e^{ i ( \alpha + \delta ) } \right], \label{eq:ma}
\eeq
where the term ``T"  and ``P" denote the ``Tree" and ``Penguin"
part of a given decay amplitude ${\cal M}$, which is proportional to
$\xi_u=V_{ub}^*V_{ud}$ or $\xi_t=V_{tb}^*V_{td}$, respectively.
While the ratio
\beq
z=\left|\frac{V_{tb}^* V_{td}}{ V_{ub}^*V_{ud}} \right| \left|\frac{P}{T}\right| \label{eq:zz}
\eeq is
proportional to the ratio of penguin (P) to tree (T) contributions,
the CKM angle $\alpha = \arg
\left[-\frac{V_{td}V_{tb}^*}{V_{ud}V_{ub}^*}\right]$ is the weak
phase, and $\delta$ is the relative strong phase between the tree
and penguin part.

Take ${\cal M}(B^+ \to K^+ \overline{K}^{*0})$ in Eq.~(\ref{eq:m3a}) as an example, its ``T" and ``P''
parts can be written as in the form of
\beq
T &=& M_{aK} C_1 + F_{aK} \left(\frac{1}{3}C_1 + C_2\right ), \label{eq:m3at}\\
P &=&  F_{eK}\left(\frac{1}{3}C_3+C_4-\frac{1}{6}C_9-\frac{1}{2}C_{10}\right)
+ F_{aK}^{P_2}\left(\frac{1}{3}C_5+C_6+\frac{1}{3}C_7+C_8 \right)\non
& & + M_{eK}\left(C_3-\frac{1}{2}C_9\right)+M_{eK}^{P_1}\left(C_5-\frac{1}{2}C_7\right)
+ M_{aK}\left(C_3+C_9\right)\non
&& + M_{aK}^{P_1}\left(C_5+C_7\right)
+ F_{aK}\left(\frac{1}{3}C_3+C_4+\frac{1}{3}C_9+C_{10}\right) .\label{eq:m3ap}
\eeq

In pQCD approach, the ratio $z$ and the strong phase $\delta$ can be calculated perturbatively.
For $B^+ \to K^+ \bar K^{*0}$ and $K^{*+} \bar K^0$ decays, for example, we find
numerically that
\beq
z(K^+ \bar K^{*0}) &=&2.1, \qquad \delta (K^+ \bar K^{*0})=-13^\circ , \non
z(K^{*+} \bar K^0) &=& 2.7 , \qquad \delta(K^{*+} \bar K^0)=-44^\circ .\label{eq:zd2}
\eeq
The major error of the ratio $z$ and the strong phase $\delta$ is induced by the uncertainty of
$\omega_b=0.4 \pm 0.04$ GeV but is small in magnitude.
The reason is that the errors induced by the uncertainties of input
parameters are largely canceled in the ratio.

From Eq.~(\ref{eq:ma}), it is easy to write the decay amplitude for the
corresponding charge conjugated decay mode
\beq
\overline{\cal M} &=& V_{ub}V_{ud}^* T -V_{tb} V_{td}^* P
= V_{ub}V_{ud}^* T \left[1 +z e^{i(-\alpha + \delta)} \right].
\label{eq:mb}
 \eeq
Therefore the CP-averaged branching ratio for $B^0 \to K K^*$ decay can be defined as
\beq
Br = (|{\cal M}|^2 +|\overline{\cal M}|^2)/2 =  \left|
V_{ub}V_{ud}^* T \right| ^2 \left[1 +2 z\cos \alpha \cos \delta +z^2
\right], \label{br}
\eeq
where the ratio $z$ and the strong phase $\delta$ have been defined in Eqs.~(\ref{eq:ma})
and (\ref{eq:zz}).

\begin{figure}[tb]
\centerline{\mbox{\epsfxsize=10cm\epsffile{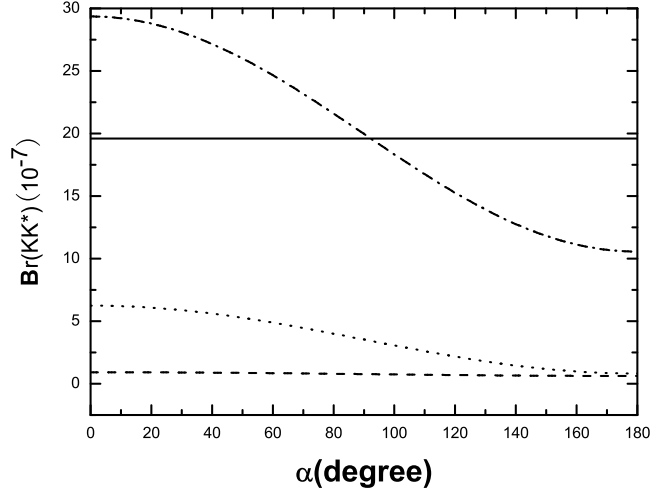}}}
 \caption{Branching ratios (in units of $10^{-7}$)
 of $B^+\to K^{*+} \ov K^0$ (dash-dotted curve), $B^+\to K^+ \ov K^{*0}$ (dotted curve),
 $B^0/\ov B^0 \to K^0 \ov K^{*0}+\ov K^0 K^{*0}$ (solid  curve),
 $B^0/\ov B^0 \to K^+ K^{*-}+ K^- K^{*+}$ (dashed
 curve)as a function of CKM angle $\alpha$.}
 \label{fig:fig5}
\end{figure}

It is a little complicate for us to calculate the branch ratios of $B^0/\bar B^0 \to f(\bar f)$,
since both $B^0$ and $\ov B^0$ can decay into the final state $f$ and $\bar f$ simultaneously.
Due to $B^0 -\ov B^0$ mixing, it is very difficult to distinguish $B^0$ from $\ov
B^0$. But it is easy to identify the final states.
Therefore we sum up $B^0/\ov B^0 \to K^0 \ov K^{*0}$ as one
channel, and $B^0/\ov B^0 \to \ov K^0 K^{*0}$ as another, although
the summed up channels are not charge conjugate states~\cite{ly}.
Similarly, we have $B^0/\ov B^0 \to K^+ K^{*-}$ as one channel, and
$B^0/\ov B^0 \to K^- K^{*+}$ as another. We show the branching
ratio of $B^0/\ov B^0 \to K^+ K^{*-}$, $B^0/\ov B^0\to K^-
K^{*+}$, $B^+\to K^+ \ov K^{*0}$ and $B^+\to K^{*+} \ov K^0$
decays as a function of $\alpha$ in Fig.~\ref{fig:fig5}.

Using  the wave functions and the input parameters as specified previously,
it is straightforward  to calculate the
branching ratios for the four considered decays. The pQCD predictions
for the branching ratios are the following
\beq
Br( B^+ \to K^+ \ov K^{*0}) &=& 3.1 ^{+1.2}_{-0.8}(\omega_b)   \times 10^{-7},
 \label{eq:brp-eta}\\
Br( B^+ \to K^{*+} \ov K^0) &=& 18.3^{+6.8}_{-4.7}( \omega_b)  \times 10^{-7},
\label{eq:brp-etap}\\
Br( B^0/\ov B^0 \to K^0 \ov K^{*0}+\ov K^0 K^{*0}) &=& 19.6
 ^{+7.9}_{-5.4}(\omega_b)\times 10^{-7}, \label{eq:br0-eta} \\
Br( B^0/\ov B^0 \to K^+ K^{*-}+ K^- K^{*+}) &=& 7.4  ^{+1.0}_{-1.3}
(\omega_b) \times 10^{-8} \label{eq:br0-etap},
\eeq
where the major error is induced by the uncertainty of $\omega_b=0.4 \pm 0.04$ GeV.

As a comparison, we also list the theoretical predictions in QCDF approach \cite{bn03b}:
\beq
Br(B^- \to K^- K^{*0}) &=& 3.0 ^{+6.0}_{-2.5} \times 10^{-7},  \label{eq:b011}\\
Br(B^- \to K^{*-} K^0) &=& 3.0 ^{+7.2}_{-2.7} \times 10^{-7},  \label{eq:b012}\\
Br(\ov B^0 \to \ov K^0 K^{*0} ) &=& 2.6 ^{+4.8}_{-2.0} \times 10^{-7}, \label{eq:b013} \\
Br(\ov B^0 \to  K^0 \ov K^{*0} ) &=& 2.9 ^{+7.3}_{-2.7} \times 10^{-7}, \label{eq:b014} \\
Br(\ov B^0 \to K^- K^{*+}) &=& 1.4 ^{+10.7}_{-1.4} \times 10^{-8} \label{eq:br0-b015},\\
Br(\ov B^0 \to  K^+ K^{*-}) &=& 1.4 ^{+10.7}_{-1.4} \times 10^{-8} \label{eq:br0-b016},
\eeq
where the individual errors as given in Refs.~\cite{bn03b} have been added in quadrature.
For $B^- \to K^- K^{*0}$ decay, the pQCD and QCDF predictions agree very well. For remaining decay modes,
the pQCD predictions are larger than the QCDF predictions by a factor of 2 to 5, although they are
still consistent with each other within errors because the theoretical uncertainties are still very large.
When compared with the experimental upper limits, the theoretical predictions
in both approaches still agree with the data.
The large differences between the pQCD and QCDF predictions will be tested by the forthcoming
precision measurements.

\subsection{CP-violating asymmetries }

Now we turn to the evaluations of the CP-violating asymmetries of $B
\to K K^*$ decays in the pQCD approach. For $B^+ \to K^+ \ov K^{*0}$
and $B^+ \to K^{*+} \ov K^0$ decays, the direct CP-violating
asymmetries $A_{CP}^{dir}$ can be defined as:
 \beq
{\cal A}_{CP}^{dir} =  \frac{|\overline{\cal M}|^2 - |{\cal M}|^2 }{
 |\overline{\cal M}|^2+|{\cal M}|^2}=
\frac{2 z \sin \alpha \sin\delta}{1+2 z\cos \alpha \cos \delta
+z^2}, \label{eq:acp1}
 \eeq
where the ratio $z$ and the strong phase $\delta$ have been
defined in previous subsection and are calculable in PQCD
approach.

\begin{figure}[tb]
\centerline{\mbox{\epsfxsize=10cm\epsffile{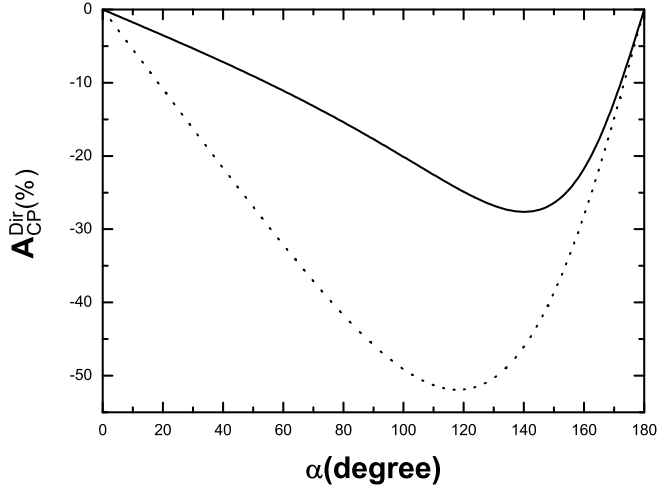}}}
\vspace{0.3cm}
 \caption{The direct CP asymmetry $A_{CP}^{dir}$ (in percentage) of
 $B^+ \to K^+ \ov K^{*0}$ (the solid curve) and $B^+ \to K^{*+} \ov
K^0$ (the dotted curve) as a function of CKM angle $\alpha$.}
 \label{fig:fig6}
\end{figure}

Using the definition in Eq.(\ref{eq:acp1}), it is easy  to calculate
the direct CP-violating asymmetries for $B^\pm \to K^\pm \ov
K^{*0}(K^{*0})$ and $B^\pm \to K^{*\pm} \ov K^0(K^0)$ decays. The
numerical results are
\beq
A_{CP}^{dir}(B^\pm \to K^\pm \ov K^{*0}(K^{*0})) &=& -0.20 \pm 0.05(\alpha)\pm 0.02(\omega_b),
\non
A_{CP}^{dir}(B^\pm \to K^{*\pm} \ov K^0(K^0)) &=& -0.49^{+0.07}_{-0.03}(\alpha) \pm 0.07(\omega_b).
\eeq
for $\alpha=100^\circ \pm 20^\circ$ and $\omega_b=0.40\pm 0.04$ GeV. These pQCD predictions
are also consistent with those in QCDF approach \cite{bn03b}:
\beq
A_{CP}^{dir}(B^\pm \to K^\pm \ov K^{*0}(K^{*0})) &=& -0.24 ^{+0.28}_{-0.39}, \non
A_{CP}^{dir}(B^\pm \to K^{*\pm} \ov K^0(K^0)) &=& -0.13 ^{+0.29}_{-0.37}, \label{eq:acp-2}.
\eeq
where the individual errors as given in Ref.~\cite{bn03b} have been added in quadrature.
In Fig.~\ref{fig:fig6}, we show the $\alpha-$dependence of the pQCD predictions of
${\cal A}_{CP}^{dir}$ for $B^\pm \to K^\pm
\ov K^{*0}(K^{*0})$ (the solid curve) and $B^\pm \to K^{*\pm} \ov
K^0(K^0)$ decay (the dotted curve), respectively.

For $B^0/\ov B^0 \to K^0 \ov K^{*0}(\ov K^0 K^{*0})$ decays, they do not exhibit CP violating
asymmetry, since they involve only penguin contributions at the leading order,
as can be seen from the decay amplitudes as given in Eqs.~(\ref{eq:m1a}) and (\ref{eq:m1b}).

We now study the CP-violating asymmetries for $B^0/\ov B^0 \to K^+
K^{*-}(K^- K^{*+})$ decays. Since both $B^0$ and $\ov B^0$ can decay to the final state
$ K^+ K^{*-}$ and $ K^{*+} K^-$, there are four decay modes.
Here we use the formulae as given in Ref.~\cite{ly}. The four time-dependent decay widths
for $B^0(t) \to K^+ K^{*-}$,$\ov B^0(t) \to K^- K^{*+}$,$B^0(t) \to
K^- K^{*+}$ and $\ov B^0(t) \to K^+ K^{*-}$
can be expressed by four basic matrix elements\cite{ly}:
\beq
g&=& \langle K^+ K^{*-} |H_{eff}| B^0\rangle,\quad
h=\langle K^+ K^{*-} |H_{eff}| \ov B^0 \rangle, \non
\ov g&=& \langle K^- K^{*+} |H_{eff}| \ov B^0\rangle ,\quad \overline h =\langle K^- K^{*+} |H_{eff}|
 B^0\rangle, \label{eq:hh}
\eeq
 which determines the decay matrix elements of $B^0 \to K^+ K^{*-}$, $\ov B^0
\to K^- K^{*+}$, $B^0 \to K^- K^{*+}$ and $\ov B^0 \to K^+ K^{*-}$
at $t=0$. The matrix elements $g$ and $\ov h$ are given in
Eqs.(\ref{eq:m2a}) and (\ref{eq:m2b}). The matrix elements $h$ and
$\ov g$ are obtained from $\ov h$ and $g$ by simple replacements of
$\xi_u \to \xi_u^*$ and $\xi_t \to \xi_t^*$: i.e., changing the sign of
the weak phases contained in the products of the CKM matrix
elements $\xi_u$ and $\xi_t$.

\begin{figure}[tb]
\centerline{\mbox{\epsfxsize=10cm\epsffile{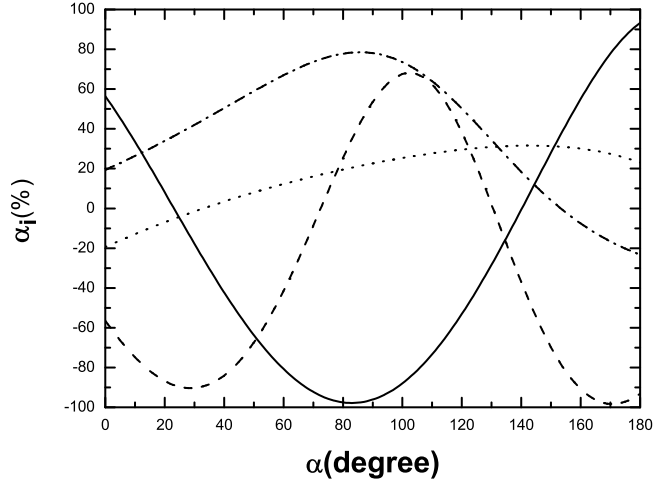}}}
\vspace{0.3cm}
 \caption{CP violating parameters of $B^0/ \ov B^0 \to K^+K^{*-}(K^-K^{*+})$ decays:
  $a_{\epsilon'}$ (dash-dotted line), $a_{\bar \epsilon'}$(dotted
  line), $a_{\epsilon+\epsilon'}$(dashed line) and
  $a_{\epsilon+\bar\epsilon'}$(solid line) as a function of CKM  angle $\alpha$}
 \label{fig:fig7}
\end{figure}

Following the general procedure,  the $B^0-\ov B^0$ mixing can be defined as
\beq
B_1=p|B^0\rangle+q| \overline B^0\rangle,\quad B_2=p|B^0\rangle-q| \overline B^0\rangle,
 \eeq
with $|p|^2 + |q|^2 = 1$. Following the notation of Ref.~\cite{ly}, the four time-dependent decay widths
of the considered decay modes can be written as
\beq
\Gamma(B^0 (t)\to K^+ K^{*-})&=&e^{-\Gamma t}\frac
 {1}{2}(|g|^2+|h|^2)\times \left \{1+ a_{\epsilon'} \cos(\Delta mt)
 +a_{\epsilon+\epsilon'}\sin(\Delta mt)\right \},\non
 \Gamma(\ov B^0 (t)\to K^+ K^{*-})&=&e^{-\Gamma t}\frac
 {1}{2}(|g|^2+|h|^2)\times \left \{1- a_{\epsilon'} \cos(\Delta mt)
 -a_{\epsilon+\epsilon'}\sin(\Delta mt)\right \},\non
 \Gamma(\ov B^0 (t)\to K^- K^{*+})&=&e^{-\Gamma t}\frac
 {1}{2}(|\ov g|^2+|\ov h|^2)\times \left \{1- a_{\ov \epsilon'} \cos(\Delta mt)
 -a_{\epsilon+\ov \epsilon'}\sin(\Delta mt)\right \}, \non
 \Gamma(B^0 (t)\to K^- K^{*+})&=&e^{-\Gamma t}\frac
 {1}{2}(|\ov g|^2+|\ov h|^2)\times \left \{1+ a_{\ov \epsilon'} \cos(\Delta mt)
 +a_{\epsilon+\ov \epsilon'}\sin(\Delta mt)\right \}, \label{eq:acpf}
 \eeq
 where the four CP violating parameters are defined as
 \beq
 a_{\epsilon'}&=&\frac{|g|^2-|h|^2}{|g|^2+|h|^2},\quad
 a_{\epsilon+\epsilon'}=\frac{-2Im(\frac{q}{p}\frac{h}{g})}{1+|h/g|^2}  \non
 a_{\ov \epsilon'}&=&\frac{|\ov h|^2-|\ov g|^2}{|\ov h|^2+|\bar g|^2},\quad
 a_{\epsilon+\bar \epsilon'}=\frac{-2Im(\frac{q}{p}\frac{\bar g}{\bar h})}{1+|\bar g/\bar h|^2},
\label{eq:aaaa} \eeq where $q/p = e^{2i\beta}$. Using the decay
amplitudes as given in Eqs.~(\ref{eq:m2a}) and (\ref{eq:m2b}), it is
straightforward to calculate the above four CP-violation parameters.
The central values of the pQCD predictions are
\beq
a_{\epsilon'}&=& 0.74, \qquad  a_{\epsilon+\epsilon'}=0.68,  \non
a_{\ov \epsilon'}&=& 0.25, \qquad  a_{\epsilon+\bar \epsilon'}=-0.88 ,
\label{eq:aaaa2}
\eeq
for $\alpha=100^\circ$. The
$\alpha-$dependence of these four CP violating parameters are shown
in Fig.~\ref{fig:fig7}. It is difficult to measure these physical
observables in current and forthcoming B meson experiments because
of its tiny branching ratio ( $ \sim 10^{-8} $ ).

At last, we will say a little more about the possible FSI effects.
As mentioned in the introduction, we here do  not consider the possible FSI effects on
the branching ratios and CP-violating asymmetries of the $B \to KK^*$ decays.
The FSI effect is in nature a subtle and complicated subject.
The smallness of FSI effects has been put forward by Bjorken \cite{b89}
based on the color transparency argument \cite{lb80}, and also supported by further
renormalization group analysis of soft gluon exchanges among initial and final state mesons \cite{soft}.
At present, the excellent agreement between the pQCD predictions for the branching ratios
and CP violating asymmetries and the precision measurements strongly support the assumption
that the FSI effects for $B\to K \pi $ decays are not important \cite{kls2001}.
For $B\to K K$ decays, fortunately, good agreement between the pQCD predictions for the branching ratios
of $B^+ \to K^+ K^0$, $B^0 \to K^+ K^- $ and $K^0 \ov{K}^0$ decays \cite{chen00a,zhu05a} and currently
available experimental measurements \cite{hfag06} indicates that the FSI effects are most possibly
not important also \cite{zhu05a}. Of course, more studies are needed about this issue, while
further consistency check between the pQCD predictions and the precision data will reveal
whether FSI effects are important or not.

\section{summary }

In this paper,  we calculate the branching ratios and CP-violating
asymmetries of $B^0/\ov B^0 \to K^0 \ov K^{*0}(\ov K^0 K^{*0})$,
$B^0/\ov B^0 \to K^+ K^{*-}(K^- K^{*+})$, $B^+ \to K^+ \ov
K^{*0}$, and $B^+ \to K^{*+} \ov K^0$ decays, together with their
charge-conjugated modes, by employing the pQCD factorization approach.

From our calculations and phenomenological analysis, we found the following results:
\begin{itemize}
\item
The pQCD predictions for the form factors of $B \to K$ and $K^*$  transitions
are
\beq
F_{0,1}^{B\to K}(0)= 0.35^{+0.06}_{-0.04}, \quad A_0^{B\to K^*} = 0.46^{+0.07}_{-0.06},
\eeq
for $\omega_b=0.40 \pm 0.04$ GeV, close to the light-cone QCD sum rule results \cite{ball05}.

\item
the pQCD predictions for the CP-averaged branching ratios are
\beq
Br( B^+ \to K^+ \ov K^{*0}) &\approx & 3.1 \times 10^{-7}, \non
Br( B^+ \to K^{*+} \ov K^0) &\approx & 18.3 \times 10^{-7}, \non
Br( B^0/\ov B^0 \to K^0 \ov K^{*0}+\ov K^0 K^{*0}) &\approx & 19.6 \times 10^{-7}, \non
Br( B^0/\ov B^0 \to K^+ K^{*-}+ K^- K^{*+}) &\approx & 7.4 \times 10^{-8}.
\eeq
The above pQCD predictions agree with the QCDF predictions within still large theoretical errors
and close to currently available experimental upper limits.

\item
For the CP-violating asymmetries of the considered decay modes, the pQCD predictions
are generally large in magnitude.

\end{itemize}

\begin{acknowledgments}

We are very grateful to Xin Liu and Hui-sheng Wang for helpful discussions.
This work is partly supported  by the National Natural
Science Foundation of China under Grant No.10275035 and 10575052,  by the Specialized
Research Fund for the doctoral Program of higher education (SRFDP) under Grant No.~20050319008,
and by the Research Foundation of Jiangsu Education Committee under Grant No.2003102TSJB137.

\end{acknowledgments}


\begin{appendix}
\section{Non-zero Transition Amplitudes}

The factorizable amplitudes $F_{eK^*}$, $F_{eK^*}^{P1}$ and $F_{eK^*}^{P2}$, $F_{aK^*}$,
$F_{aK^*}^{P1}$ and $F_{aK^*}^{P2}$ have been given in Sec.~{\ref{sec:p-c}}. The remaining
factorizable transition amplitudes in $B \to K K^* $ decays are written as:
\beq
F_{eK} &=& 4 \sqrt{2} \pi G_F C_F f_{K^*} m_B^4
           \int_{0}^{1}d x_{1}d x_{3}\,\int_{0}^{\infty} b_1d b_1 b_3d b_3\,\phi_B(x_1,b_1) \non
     & &   \cdot \left\{ \left[(1+x_3)\phi_K^A(x_3,b_3)+r_K  (1-2x_3) \left(\phi_K^P(x_3,b_3) +\phi_K^T(x_3,b_3)
           \right)  \right] \right.  \non
     & &   \left. \cdot \alpha_s (t_e^1)h_e(x_1,x_3,b_1,b_3)\exp[-S_{a}(t_e^1)]\right. \non
     & &   \left. + 2 r_K \phi_K^P(x_3,b_3) \alpha_s (t_e^2)
           h_e(x_3,x_1,b_3,b_1)\exp[-S_{a}(t_e^2)] \right \}, \label{eq:fek}\\
F_{eK}^{P_1} &=& F_{eK}, \\
F_{aK} &=& 4 \sqrt{2} \pi G_F C_F f_B m_B^4
          \int_{0}^{1}dx_{2}\,d x_{3}\,\int_{0}^{\infty} b_2d b_2b_3d b_3 \, \non
     &&   \cdot \left\{ \left[ x_3 \phi_{K^*}(x_3,b_3) \phi_K^A(x_2,b_2)
          +2 r_{K^*} r_K \phi_K^P(x_2,b_2) \left((1+x_3)\phi_{K^*}^s(x_3,b_3)\right.\right.\right. \non
     &&   \left.\left.\left.-(1-x_3) \phi_{K^*}^t(x_3,b_2) \right)\right]\alpha_s(t_e^3)
          h_a(x_2,x_3,b_2,b_3)\exp[-S_{d}(t_e^3)]\right. \non
     &&   \left. -\left[x_2 \phi_{K^*}(x_3,b_3) \phi_K^A(x_2,b_2)+2 r_{K^*}r_K \phi_{K^*}^s(x_3,b_3)
          \left((1+x_2)\phi_K^P(x_2,b_2)\right.\right.\right. \non
     &&   \left.\left.\left.-(1-x_2)\phi_K^T(x_2,b_2) \right) \right] \alpha_s(t_e^4)
          h_a(x_3,x_2,b_3,b_2)\exp[-S_{d}(t_e^4)]\right \}, \label{eq:faksb}\\
F_{aK}^{P_1} &=& -F_{aK},\\
F_{a{K}}^{P_2} &=& 8 \sqrt{2} G_F \pi C_F  m_B^4 f_B
          \int_{0}^{1}d x_{2}\,dx_{3}\,\int_{0}^{\infty} b_2d b_2b_3d b_3 \, \non
     &&   \cdot \left\{ \left[2 r_K \phi_{K^*}(x_3, b_3) \phi_K^P(x_2,b_2)
          + x_3 r_{K^*} \left(\phi_{K^*}^s(x_3, b_3)- \phi_{K^*}^t(x_3,b_2)\right)
          \phi_K^A(x_2,b_2) \right]\right. \non
     &&   \left.\cdot \alpha_s(t_e^3)h_a(x_2,x_3,b_2,b_3)\exp[-S_{d}(t_e^3)]\right. \non
     &&   \left.+\left[2 r_{K^*}\phi_{K^*}^s(x_3,b_3) \phi_K^A(x_2,b_2)+ x_2 r_K
          \left(\phi_K^P(x_2,b_2)-\phi_K^T(x_2,b_2)\right)\phi_{K^*}(x_3,b_3)\right]\right.  \non
     &&   \left. \cdot  \alpha_s(t_e^4) h_a(x_3,x_2,b_3,b_2)\exp[-S_{d}(t_e^4)]\right \}\;. \label{eq:faksp2b}
\eeq

For $B \to K K^* $ decays, the non-factorizable transition
amplitudes not shown explicitly in Sec.~\ref{sec:p-c} are written as:
\beq
M_{eK}  &=&- \frac{16}{\sqrt{3}} G_F \pi C_F  m_B^4
             \int_{0}^{1}d x_{1}d x_{2}\,dx_{3}\,\int_{0}^{\infty} b_1d b_1 b_2d b_2\, \phi_B(x_1,b_1)
             \phi_{K^*}(x_2,b_2)  \non
        & &  \cdot \left\{ \left[-x_2\phi_K^A(x_3, b_2)+ r_Kx_3\left(\phi_K^P(x_3,b_2)
             -\phi_K^T(x_3,b_2)\right)\right] \right.  \non
        & &  \left.\cdot \alpha_s(t_f^1) h_f^1(x_1,x_2,x_3,b_1,b_2)\exp[-S_{b}(t_f^1)] \right. \non
        & &  \left. -\left[(x_2-x_3-1)\phi_K^A(x_3, b_2)+ r_Kx_3\left(\phi_K^P(x_3,b_2)
             +\phi_K^T(x_3,b_2)\right)\right] \right. \non
        & &  \left.\cdot \alpha_s(t_f^2) h_f^2(x_1,x_2,x_3,b_1,b_2)\exp[-S_{b}(t_f^2)] \right\},\label{eq:mek}
\eeq
\beq
M_{eK}^{P_1} &=& -\frac{16} {\sqrt{3}} G_F \pi C_F r_{K^*} m_B^4
                 \int_{0}^{1}d x_{1}d x_{2}\,d x_{3}\,\int_{0}^{\infty} b_1d b_1 b_2db_2\, \phi_B(x_1,b_1) \non
        & &  \cdot \left\{ \left[ x_2 \phi_K^A(x_3,b_2)\left(\phi_{K^*}^s(x_2,b_2)-\phi_{K^*}^t(x_2,b_2)\right)
             - r_K\left(x_1\left(\phi_K^P(x_3,b_2) \right.\right.\right.\right. \non
        &&   \left.\left.\left.\left. - \phi_K^T(x_3,b_2)\right)\left(\phi_{K^*}^s(x_2,b_2)
             -\phi_{K^*}^t(x_2,b_2)\right)-x_2\left(\phi_K^P(x_3,b_2)\right.\right.\right.\right. \non
        &&   \left.\left.\left.\left. -\phi_K^T(x_3,b_2)\right) \left(\phi_{K^*}^s(x_2,b_2)
             -\phi_{K^*}^t(x_2,b_2)\right) -x_3\left(\phi_K^P(x_3,b_2)+\phi_K^T(x_3,b_2)\right)
             \right.\right.\right. \non
        &&  \left.\left.\left.\cdot \left(\phi_{K^*}^s(x_2,b_2)+\phi_{K^*}^t(x_2,b_2)\right)\right)\right]
            \alpha_s(t_f^1)h_f^1(x_1,x_2,x_3,b_1,b_2)\exp[-S_{b}(t_f^1)]\right. \non
        &&  \left.+\left[ (1-x_2) \phi_K^A(x_3,b_2) \left(\phi_{K^*}^s(x_2,b_2)-\phi_{K^*}^t(x_2,b_2)\right)
            - r_K\left(x_1\left(\phi_K^P(x_3,b_2) \right.\right.\right.\right. \non
        &&  \left.\left.\left.\left. - \phi_K^T(x_3,b_2)\right)\left(\phi_{K^*}^s(x_2,b_2)
            -\phi_{K^*}^t(x_2,b_2)\right)-(1-x_2)\left(\phi_K^P(x_3,b_2)\right.\right.\right.\right. \non
        &&  \left.\left.\left.\left. -\phi_K^T(x_3,b_2)\right) \left(\phi_{K^*}^s(x_2,b_2)
            -\phi_{K^*}^t(x_2,b_2)\right) -x_3\left(\phi_K^P(x_3,b_2)+\phi_K^T(x_3,b_2)\right)
            \right.\right.\right. \non
        &&  \left.\left.\left.\cdot \left(\phi_{K^*}^s(x_2,b_2)+\phi_{K^*}^t(x_2,b_2)\right)\right)\right]
            \alpha_s(t_f^2)h_f^2(x_1,x_2,x_3,b_1,b_2)\exp[-S_{b}(t_f^2)] \right \}, \label{eq:mekp1}
\eeq
\beq
M_{aK}  &=& -\frac{16}{\sqrt{3}} \pi G_F C_F  m_B^4 \int_{0}^{1}d x_{1}d x_{2}\,dx_{3}\,\int_{0}^{\infty}
            b_1d b_1 b_2d b_2\, \phi_B(x_1,b_1) \non
        & &  \cdot \left\{\left[x_2 \phi_{K^*}(x_2,b_2)\phi_K^A(x_3, b_2)
             +r_{K^*} r_K\left(\left(\left(x_2+x_3+2\right)\phi_{K^*}^s(x_2,b_2)\right.\right.\right.\right. \non
        & &  \left.\left.\left.\left. +\left(x_2-x_3\right)\phi_{K^*}^t(x_2,b_2)\right)\phi_K^P(x_3,b_2)
             +\phi_K^T(x_3,b_2) \left(-x_3\left(\phi_{K^*}^s(x_2,b_2)\right.\right.\right.\right.\right.  \non
        & &  \left.\left.\left.\left.\left. -\phi_{K^*}^t(x_2,b_2)\right)-2\phi_{K^*}^t(x_2,b_2)
             +x_2\left(\phi_{K^*}^s(x_2,b_2) +\phi_{K^*}^t(x_2,b_2)\right)\right)\right)\right]\right. \non
        & &  \cdot \left.\alpha_s(t_f^3)h_f^3(x_1,x_2,x_3,b_1,b_2)\exp[-S_{c}(t_f^3)] \right.  \non
        & &  \left. - \left[x_3 \phi_{K^*}(x_2,b_2)\phi_K^A(x_3,b_2)+r_{K^*}r_K\left(x_2\left(\phi_K^P(x_3,b_2)
             -\phi_K^T(x_3,b_2)\right)\right.\right.\right. \non
        & & \cdot\left.\left.\left.\left(\phi_{K^*}^s(x_2,b_2)-\phi_{K^*}^t(x_2,b_2)\right)
            +x_3\left(\phi_K^P(x_3,b_2)+\phi_K^T(x_3,b_2)\right) \right.\right.\right. \non
        & & \cdot\left.\left.\left.\left(\phi_{K^*}^s(x_2,b_2)+\phi_{K^*}^t(x_2,b_2)\right)\right)\right]
            \alpha_s(t_f^4)h_f^4(x_1,x_2,x_3,b_1,b_2)\exp[-S_{c}(t_f^4)]  \right\}, \label{eq:mak}\\
M_{a{K}}^{P_1}  &=& M_{a{K^*}}^{P_1}, \\
M_{a{K}}^{P_2}  &=& M_{a{K^*}}^{P_2},
\eeq
where $r_K= m_0^K/m_B $ with $m_0^K=m_K^2/(m_s+m_d)$.

\section{Related functions}

We show here the function $h_i$'s, coming from the Fourier
transformations  of $H^{(0)}$,
 \beq
 h_e(x_1,x_3,b_1,b_3)&=&
 K_{0}\left(\sqrt{x_1 x_3} m_B b_1\right)
 \left[\theta(b_1-b_3)K_0\left(\sqrt{x_3} m_B
b_1\right)I_0\left(\sqrt{x_3} m_B b_3\right)\right.
 \non
& &\;\left. +\theta(b_3-b_1)K_0\left(\sqrt{x_3}  m_B b_3\right)
I_0\left(\sqrt{x_3}  m_B b_1\right)\right] S_t(x_3), \label{he1}
\eeq
 \beq
 h_a(x_2,x_3,b_2,b_3)&=&
 K_{0}\left(i \sqrt{x_2 x_3} m_B b_2\right)
 \left[\theta(b_3-b_2)K_0\left(i \sqrt{x_3} m_B
b_3\right)I_0\left(i \sqrt{x_3} m_B b_2\right)\right.
 \non
& &\;\;\;\;\left. +\theta(b_2-b_3)K_0\left(i \sqrt{x_3}  m_B
b_2\right) I_0\left(i \sqrt{x_3}  m_B b_3\right)\right] S_t(x_3),
\label{he3}
\eeq
 \beq
 h_{f}^{(j)}(x_1,x_2,x_3,b_1,b_2) &=&
 \biggl\{\theta(b_2-b_1) \mathrm{I}_0(M_B\sqrt{x_1 x_3} b_1)
 \mathrm{K}_0(M_B\sqrt{x_1 x_3} b_2)
 \non
&+ & (b_1 \leftrightarrow b_2) \biggr\}  \cdot\left(
\begin{matrix}
 \mathrm{K}_0(M_B D_{(j)} b_2), & \text{for}\quad D^2_{(j)}>0 \\
 \frac{\pi i}{2} \mathrm{H}_0^{(1)}(M_B\sqrt{|D^2_{(j)}|}\ b_2), &
 \text{for}\quad D^2_{(j)}<0
\end{matrix}\right), \label{eq:pp1}
 \eeq
 \beq
h_f^3(x_1,x_2,x_3,b_1,b_2) &=& \biggl\{\theta(b_1-b_2)
\mathrm{K}_0(i \sqrt{x_2 x_3} b_1 M_B)
 \mathrm{I}_0(i \sqrt{x_2 x_3} b_2 M_B)+(b_1 \leftrightarrow b_2) \biggr\}
 \non
& & \cdot
 \frac{\pi i}{2} \mathrm{H}_0^{(1)}(\sqrt{x_1+x_2+x_3-x_1 x_3-x_2 x_3}\ b_1 M_B),
\label{eq:pp3}
\eeq
\beq h_f^4(x_1,x_2,x_3,b_1,b_2) &=&
 \biggl\{\theta(b_1-b_2) \mathrm{K}_0(i \sqrt{x_2 x_3} b_1 M_B)
 \mathrm{I}_0(i \sqrt{x_2 x_3} b_2 M_B) \non
&+& (b_1 \leftrightarrow b_2) \biggr\} \cdot \left(
\begin{matrix}
 \mathrm{K}_0(M_B F_{(1)} b_1), & \text{for}\quad F^2_{(1)}>0 \\
 \frac{\pi i}{2} \mathrm{H}_0^{(1)}(M_B\sqrt{|F^2_{(1)}|}\ b_1), &
 \text{for}\quad F^2_{(1)}<0
\end{matrix}\right),\label{eq:pp4}
\eeq
where j=1 and 2, $J_0$ is the Bessel function
and $K_0$, $I_0$ are modified Bessel functions $K_0 (-i x) =
-(\pi/2) Y_0 (x) + i (\pi/2) J_0 (x)$, and $F^2_{(1)}$ , $D_{(j)}$'s
are defined by
\beq
F^2_{(1)}&=&(x_1 -x_2) x_3\;,\non
D^2_{(1)}&=&(x_1-x_2) x_3\;,\non
D^2_{(2)}&=&-(1-x_1-x_2) x_3\; .
\eeq

The threshold resummation form factor $S_t(x_i)$ is adopted from
Ref.\cite{kurimoto}
\beq
S_t(x)=\frac{2^{1+2c} \Gamma (3/2+c)}{\sqrt{\pi} \Gamma(1+c)}[x(1-x)]^c,
\eeq
where the parameter $c=0.3$. This function is normalized to unity.

The Sudakov factors used in the text are defined as
\beq
S_{a}(t) &=& s\left(x_1 m_B/\sqrt{2}, b_1\right) +s\left(x_3 m_B/\sqrt{2},
b_3\right) +s\left((1-x_3) m_B/\sqrt{2}, b_3\right) \non
&&-\frac{1}{\beta_1}\left[\ln\frac{\ln(t/\Lambda)}{-\ln(b_1\Lambda)}
+\ln\frac{\ln(t/\Lambda)}{-\ln(b_3\Lambda)}\right],
\label{wp}\\
S_{b}(t) &=& s\left(x_1 m_B/\sqrt{2}, b_1\right)
 +s\left(x_2 m_B/\sqrt{2}, b_2\right)
+s\left((1-x_2) m_B/\sqrt{2}, b_2\right) \non
 && +s\left(x_3
m_B/\sqrt{2}, b_1\right) +s\left((1-x_3) m_B/\sqrt{2}, b_1\right)
\non
 & &-\frac{1}{\beta_1}\left[2
\ln\frac{\ln(t/\Lambda)}{-\ln(b_1\Lambda)}
+\ln\frac{\ln(t/\Lambda)}{-\ln(b_2\Lambda)}\right],
\label{Sc}\\
S_{c}(t) &=& s\left(x_1 m_B/\sqrt{2}, b_1\right)
 +s\left(x_2 m_B/\sqrt{2}, b_2\right)
+s\left((1-x_2) m_B/\sqrt{2}, b_2\right) \non
 && +s\left(x_3
m_B/\sqrt{2}, b_2\right) +s\left((1-x_3) m_B/\sqrt{2}, b_2\right)
\non
 &
&-\frac{1}{\beta_1}\left[\ln\frac{\ln(t/\Lambda)}{-\ln(b_1\Lambda)}
+2\ln\frac{\ln(t/\Lambda)}{-\ln(b_2\Lambda)}\right],
\label{Se}\\
S_{d}(t) &=& s\left(x_2 m_B/\sqrt{2}, b_2\right)
 +s\left(x_3 m_B/\sqrt{2}, b_3\right)
+s\left((1-x_2) m_B/\sqrt{2}, b_2\right) \non
 &+& s\left((1-x_3) m_B/\sqrt{2}, b_3\right)
-\frac{1}{\beta_1}\left[\ln\frac{\ln(t/\Lambda)}{-\ln(b_2\Lambda)}
+\ln\frac{\ln(t/\Lambda)}{-\ln(b_3\Lambda)}\right], \label{ww}
\eeq
where the function $s(q,b)$ are defined in the Appendix A of
Ref.\cite{luy01}. The scale $t_i$'s in the above equations are
chosen as
\beq
t_{e}^1 &=& {\rm max}(\sqrt{x_3} m_B,1/b_1,1/b_3)\;,\non
t_{e}^2 &=& {\rm max}(\sqrt{x_1}m_B,1/b_1,1/b_3)\;,\non
t_{e}^3 &=& {\rm max}(\sqrt{x_3}m_B,1/b_2,1/b_3)\;,\non
t_{e}^4 &=& {\rm max}(\sqrt{x_2}m_B,1/b_2,1/b_3)\;,\non
t_{f}^1 &=& {\rm max}(\sqrt{x_1 x_3}m_B,
\sqrt{(x_1-x_2) x_3} m_B,1/b_1,1/b_2)\;,\non
t_{f}^2 &=& {\rm max}(\sqrt{x_1 x_3}m_B,
\sqrt{(1-x_1-x_2) x_3} m_B,1/b_1,1/b_2)\;,\non
t_{f}^3 &=& {\rm max}(\sqrt{x_1+x_2+x_3-x_1 x_3-x_2 x_3}m_B,
\sqrt{x_2 x_3} m_B,1/b_1,1/b_2)\;,\non
 t_{f}^4 &=& {\rm max}(\sqrt{(x_1-x_2) x_3} m_B,\sqrt{x_2 x_3} m_B,1/b_1,1/b_2)\;.
\eeq

\end{appendix}


\end{document}